\documentclass[sigconf]{acmart}

\usepackage[utf8]{inputenc}
\usepackage[T1]{fontenc}
\usepackage{graphicx}

\usepackage{booktabs}
\usepackage{multirow}
\usepackage{array}
\usepackage{tabularx}
\usepackage{rotating}

\usepackage[dvipsnames,table]{xcolor}

\usepackage[normalem]{ulem}
\usepackage{colortbl}

\usepackage{balance}
\usepackage{microtype}
\usepackage{framed}

\usepackage{multicol}
\usepackage{verbatim}
\usepackage{comment}

\usepackage[ampersand]{easylist}
\usepackage[inline]{enumitem}

\usepackage[roman]{parnotes}
\usepackage{footmisc}
\usepackage{ccicons}
\usepackage{textcomp}
\usepackage{mathtools}
\usepackage{algorithmicx}
\usepackage{tikz}
\usepackage{hyphenat}

\PassOptionsToPackage{hyphens}{url}

\AtBeginDocument{%
  \providecommand\BibTeX{{%
    \normalfont B\kern-0.5em{\scshape i\kern-0.25em b}\kern-0.8em\TeX}}}

\def\tablename{Table}

\DeclareTextFontCommand{\mytexttt}{\ttfamily\hyphenchar\font=45\relax}

\newcolumntype{L}[1]{>{\hsize=.8\hsize\raggedright\arraybackslash}m{#1}}
\newcolumntype{R}[1]{>{\hsize=.8\hsize\raggedleft\arraybackslash}m{#1}}
\newcolumntype{C}[1]{>{\hsize=.8\hsize\centering\arraybackslash}m{#1}}

\definecolor{navy}{rgb}{0.0, 0.0, 0.5}

\newcommand{\takeaway}[1]{\textcolor{navy}{\textbf{\emph{#1}}}}

\usepackage[most]{tcolorbox}

\usepackage[dvipsnames,table]{xcolor} 
\usepackage{soul}                     
\sethlcolor{white}                   

\soulregister\cite7
\soulregister\citep7
\soulregister\citet7
\soulregister\Cite7
\soulregister\Citep7
\soulregister\Citet7
\soulregister\ref7
\soulregister\eqref7
\soulregister\autoref7
\soulregister\footnote7

\newtcolorbox{repobox}{%
  colback=black!3,
  colframe=black!65,
  boxrule=0.5pt,
  arc=2mm,
  left=4mm,right=4mm,top=2mm,bottom=2mm,
  enhanced,
  borderline west={2pt}{0pt}{black!65}}

\newif\ifdraft
\draftfalse 

\setlist[itemize]{noitemsep, topsep=0pt}

\copyrightyear{2026}
\acmYear{2026}
\setcopyright{cc}
\setcctype{by}
\acmConference[ICSE '26]{2026 IEEE/ACM 48th International Conference on Software Engineering}{April 12--18, 2026}{Rio de Janeiro, Brazil}
\acmBooktitle{2026 IEEE/ACM 48th International Conference on Software Engineering (ICSE '26), April 12--18, 2026, Rio de Janeiro, Brazil}
\acmPrice{}
\acmDOI{10.1145/3744916.3787782}
\acmISBN{979-8-4007-2025-3/2026/04}

\ccsdesc[500]{Human-centered computing~Open source software}
\ccsdesc[500]{Software and its engineering~Collaboration in software development}

\begin{document}

\title{An Empirical Analysis of Community and Coding Patterns in OSS4SG vs. Conventional OSS}

\author{Mohamed Ouf, Shayan Noei, Zeph Van Iterson, Mariam Guizani, Ying Zou}
\affiliation{%
  \institution{Queen's University, Canada}
  \city{}
  \country{}
}

\email{{24blr2,s.noei,20zrvi,mariam.guizani,ying.zou}@queensu.ca}

\begin{abstract}
Open Source Software for Social Good (OSS4SG) projects aim to address critical societal challenges, such as healthcare access and community safety. Understanding the community dynamics and contributor patterns in these projects is essential for ensuring their sustainability and long-term impact. However, while extensive research has focused on conventional Open Source Software (OSS), little is known about how the mission-driven nature of OSS4SG influences its development practices. To address this gap, we conduct a large-scale empirical study of 1,039 GitHub repositories, comprising 422 OSS4SG and 617 conventional OSS projects, to compare community structure, contributor engagement, and coding practices. Our findings reveal that OSS4SG projects foster significantly more stable and ``sticky'' (63.4\%) communities, whereas conventional OSS projects are more ``magnetic'' (75.4\%), attracting a high turnover of contributors. OSS4SG projects also demonstrate consistent engagement throughout the year, while conventional OSS communities exhibit seasonal fluctuations. Additionally, OSS4SG projects rely heavily on core contributors for both code quality and issue resolution, while conventional OSS projects leverage casual contributors for issue resolution, with core contributors focusing primarily on code quality.
\end{abstract}

\keywords{open source, social good, code quality, community dynamics, software engineering}

\maketitle


\section{Introduction}
\label{sec:intro}

Open Source Software (OSS) plays a critical role in software development as well as in workforce development, where contributors join projects to learn ~\cite{gerosa2021motivation}, showcase their technical skills~\cite{Marlow2013CSCW}, and improve their career paths ~\cite{Singer2013CSCW, trinkenreich2020hidden}
Although OSS formally emerged as a distinct movement several decades ago, its roots lie in a vibrant community of hobbyists and volunteers who have long shared and collaborated on software. Over time, this volunteer-driven model evolved into a mainstream development paradigm that combines paid and unpaid contributors. As demonstrated by Guizani et al. \cite{guizani2023rules}, companies, from startups to tech giants, have come to view OSS as a strategic asset, actively contributing to its development as a means of maintaining competitive advantage and building verifiable trust. 


While much of the research on Open Source Software (OSS) has focused on understanding its community dynamics and development processes as a whole, there is an emerging recognition that not all OSS projects share the same underlying motivations. In particular, Open Source Software for Social Good (OSS4SG) projects, driven by a focus on societal impact rather than technical advancement or career growth \cite{Huang2021}, may involve different collaborative dynamics and development processes. OSS4SG projects are designed to address critical societal issues such as healthcare access in remote areas or safety for marginalized groups. For example, \textsc{CommCare} \cite{commcare} helps health workers track disease outbreaks in developing nations, and \textsc{Little Window} \cite{little-window} provides resources to support domestic violence victims in escaping abusive relationships. In this study, we explore the hypothesis that OSS4SG and conventional OSS projects exhibit distinct community dynamics and development practices, presenting the first empirical analysis of community and coding patterns in OSS4SG compared to conventional OSS.

\hl{While prior research has examined OSS4SG through qualitative approaches \cite{fang2023four, Huang2021}, we empirically assess whether the social missions of OSS4SG translate into distinct development practices through quantitative analysis of 289 repositories (i.e., 198 OSS4SG and 91 conventional OSS) spanning over three million commits} Our analysis investigates (1) project and community characteristics, including project stability and community retention, (2) contributor engagement across project life stages, and (3) code quality and maintainability through static analysis and issue resolution velocity.

We frame our investigation around the following questions: \begin{enumerate}[leftmargin=*]
    \item[] \textbf{RQ1. }\textit{How do project characteristics, stability, contributor retention differ between OSS4SG and conventional OSS projects?}
    
    \vspace{2mm}
    
    \item [] While conventional OSS projects are significantly more popular, OSS4SG projects enable significantly deeper core contributor engagement resulting in higher responsiveness and more consistent contributor retention. The majority of OSS4SG projects form \emph{sticky} communities (63.4\%, stable and attractive), whereas the majority of OSS projects are \emph{magnetic} (75.4\%, attractive or unstable).

    \vspace{2mm}
    
    \item[] \textbf{RQ2. }\textit{How do contributor engagement patterns cluster and evolve differently between conventional OSS and OSS4SG?}
    
    \vspace{2mm}
    
    \item[] We found that OSS4SG projects exhibit significantly higher and more consistent year-round activity among core members, especially in early stages. In contrast, conventional OSS projects face seasonal engagement and diminishing core involvement.

    \vspace{2mm}
    
    \item[] \textbf{RQ3. }\textit{How do contributor engagement patterns affect code quality in OSS4SG and conventional OSS?}

    \vspace{2mm}
    
    \item [] Contributor engagement patterns shape quality outcomes differently. OSS4SG relies on core contributors for both structural integrity and issue resolution, with quality deteriorating as core activity declines. In contrast, conventional OSS projects rely on a distributed approach, relying on casual contributors for issue resolution and core members for managing complexity.

\end{enumerate}





Our research contributes the following. 

\begin{enumerate}
\item We conduct the first large-scale empirical study on 422 OSS4SG and 617 conventional OSS projects (>3M commits) to explore how mission-driven goals impact community dynamics, engagement, and code quality.
\item We identify two sustainability models: OSS4SG projects foster \emph{sticky} communities, while conventional OSS projects create \emph{magnetic} communities.
\item We demonstrate that mission-driven goals influence engagement patterns, with OSS4SG projects attracting long-term contributors, enhancing project quality and issue resolution.
\item We offer insights and actionable recommendations for both OSS4SG and conventional OSS maintainers.
\item \hl{We provide an open replication package, including the dataset and analysis scripts, and supplemental material}: {\url{ https://doi.org/10.5281/zenodo.16337983}}.
\end{enumerate}

\section{Related Work}
\label{sec:related}

A sustainable OSS ecosystem is one that maintains a healthy influx of contributors, manageable turnover, and a healthy code base. All these three aspects are interconnected. An over-reliance on a limited set of contributors can hinder maintenance and increase the risk of turnover. High turnover signals challenges in attracting and retaining contributors \cite{feng2022case, guizani2022perceptions, goggins2021making} as well as maintaining project momentum~\cite{goggins2021making}. \citet{steinmacher2015social,Steinmacher.Chaves.ea_2014} identified 58 barriers faced by newcomers and analyzed how the answers to newcomers' first emails influenced their likelihood to stay \cite{steinmacher2013newcomers}. \citet{guizani2021long} developed a conceptual model of the ongoing challenges faced by OSS contributors, revealing that several issues, such as poor code quality and delays in contribution review or acceptance, persist well beyond the onboarding phase. \citet{pinto2016more} found that nearly half of a project's contributors submitted a single contribution and never returned, with many never having a single pull request (PR) accepted~\cite{steinmacher2018almost}.
Our work builds on these findings by examining how contributor retention, project stability, and code quality differ between OSS4SG and conventional OSS projects through large-scale empirical analysis.

High turnover negatively affects team cognition, performance, and code quality, increasing bug density and delaying issue resolution~\cite{ferreira2020turnover, miller2019people, sharma2012examining, foucault2015impact, nassif2017revisiting}. For example, \citet{ferreira2020turnover} showed that projects with unstable core teams take nearly twice as long to fix bugs, highlighting how developer attrition undermines maintainability. Attracting and retaining contributors involves addressing both personal motivations and project-level practices. While intrinsic motivations such as skill development and ideology remain important~\cite{Hars2002}, recent work shows growing emphasis on social incentives like altruism and recognition~\cite{gerosa2021shifting}. Identifying clear goals and formally acknowledging diverse contributions can improve both recruitment and retention~\cite{guizani2022attracting, zhou2012make, trinkenreich2020hidden}. We extend this research by examining how different contributor engagement patterns, beyond just turnover, affect code quality and issue resolution in both OSS4SG and conventional OSS projects.

The growing emphasis on social and altruistic motivations parallels Open Source Software for Social Good (OSS4SG) projects, a distinct category within the broader OSS ecosystem that is driven by social impact goals~\cite{Huang2021}. OSS4SG projects explicitly address societal challenges such as healthcare access, climate resilience, and community safety~\cite{Huang2021}. ~\citet{Huang2021} conducted 21 semi-structured interviews with OSS4SG contributors, followed by a large-scale survey with 517 respondents from both OSS4SG and general OSS communities. Their findings show that OSS4SG contributors emphasize societal impact over personal career benefits, indicating notable motivational differences. 
Building upon this, \citet {fang2023four} conducted a longitudinal study spanning four years, involving 1,361 students contributing to 443 OSS projects. They compared contributions to OSS4SG projects against general OSS in terms of student motivation, project selection, and community reception. Their analysis revealed that OSS4SG projects exhibited significantly higher pull request acceptance rates. The study finds that adding a lightweight educational module highlighting OSS4SG projects, successfully increased student awareness and contributions toward OSS4SG initiatives \cite{fang2023four}.

Prior research has established that OSS4SG contributors are driven by distinct motivations centered on societal impact rather than personal career advancement. \citet{Huang2021} have documented these motivational differences through interviews and surveys with contributors, while \citet{fang2023four} demonstrate higher pull request acceptance rates in a four-year longitudinal study of student contributions to OSS4SG projects. 
\hl{While prior studies have established why contributors are attracted to OSS4SG, they did not analyze how these motivations shape development behaviors and project outcomes.} Our study complements this prior work by conducting a large-scale quantitative analysis of over a thousand repositories and millions of commits drawn from software repositories. \hl{We examine whether mission-driven motivations translate into measurably different patterns in project stability, contributor retention across project lifecycles, and code quality, revealing how social mission shapes OSS4SG community dynamics, contributor engagement, and code quality.}

\section{Dataset Construction and Metrics}
\label{sec:method}

This section details our dataset construction. To build the dataset of our study, we use a recent dataset ~\cite{Huang2021dataset}. \hl{At the time of our data collection, 617 conventional OSS and 422 OSS4SG projects were publicly accessible, as 40 projects became unavailable due to repository deletion or privacy changes.}
We follow a filtering mechanism similar to previous work by \citet{Pantiuchina2021} to ensure the quality and activity status of our dataset. We include projects that: have at least (1) 10 contributors, (2) 500 commits (3)  50 closed Pull Requests, and project that (4) have more than one year of history, and (5) were updated at least once in the last year. \hl{As a result, our final dataset includes a total of 91 OSS and 198 OSS4SG projects. And this filtering process yields comparable groups of OSS and OSS4SG projects by removing toy projects, inactive projects and project with insufficient development history while maintaining the statistical characteristics for representativeness of the original dataset with 95\% confidence level and 5\% margin of error \cite{israel1992determining, ahmad2017determining}} Table~\ref{tab:filter-steps} details the outcome of each filtering step.

\begin{table}[!htbp]
    \centering
    \caption{Number of Projects Removed by Each Filtering Criterion.}
    \label{tab:filter-steps}
    \begin{tabular}{l|c|c|c}
    \hline
    \textbf{Filtering Step} & \textbf{\# Projects} & \textbf{\# OSS4SG} & \textbf{\# OSS} \\ \hline
    Initial Dataset \cite{Huang2021} & 1,039 & 422 & 617 \\ \hline
    Removed: < 10 Contributors & 517 & 122 & 395 \\
    Removed: < 500 Commits & 131 & 37 & 94 \\
    Removed: < 50 Closed PRs & 21 & 9 & 12 \\
    Removed: Inactive > 1 Year & 81 & 56 & 25 \\
    Removed: Lifespan < 1 Year & 0 & 0 & 0 \\ \hline
    \textbf{Remaining} & \textbf{289} & \textbf{198} & \textbf{91} \\ \hline
    \end{tabular}
\end{table}

To address our research questions, we combined data from the GitHub REST API~\cite{githubAPI}, the GitHub GraphQL API~\cite{githubGraphQL}, and locally cloned repositories. The REST API provided project-level attributes such as stars, forks, issues, and pull requests, while the GraphQL API retrieved contributor-specific data that the REST API does not fully expose. The locally cloned repositories reduced the mining time to obtain all commits. For each project, we used Git's reset functionality to create snapshots at specific timestamps corresponding to different project life stages, allowing us to analyze how code quality and contributor patterns evolved over time. \hl{We perform a correlation analysis to identify highly correlated Metrics. Since our data do not follow a normal distribution, we apply Spearman rank hierarchical clustering \cite{spearman1961general} with a correlation threshold of 0.7 \cite{ghaleb2022popularity}.} Table~\ref{tab:raw-rq-metrics-clustered} summarizes the metrics collected for all research questions.

\section{Project characteristics, Stability, and Retention (RQ1)}
\subsection{Motivation}

Prior studies establish that OSS4SG projects attract contributors driven by social impact, unlike conventional OSS where contributors seek personal benefits in a ``scratch-your-own-itch''~\cite{Raymond1999} fashion or career advancement  ~\cite{Huang2021}. If OSS4SG projects attract people with different motivations, does this difference translate to into difference in overall project characteristics, stability and contributor retention? A systematic, data-driven comparison is needed to identify these differences, enabling each ecosystem to adopt practices suited to their unique contributor base and learn from each other.

\subsection{Approach}
\newcommand{\hlpurple}[1]{{\sethlcolor{violet!25}\hl{#1}}}  
\newcommand{\hlcyan}[1]{{\sethlcolor{cyan!25}\hl{#1}}}      
\newcommand{\hlorange}[1]{{\sethlcolor{orange!25}\hl{#1}}}  
\newcommand{\hlgreen}[1]{{\sethlcolor{green!25}\hl{#1}}}    
\newcommand{\hlred}[1]{{\sethlcolor{gray!30}\hl{#1}}}      
\newcommand{\hlblue}[1]{{\sethlcolor{yellow!30}\hl{#1}}}    
\newcommand{\hlmagenta}[1]{{\sethlcolor{teal!25}\hl{#1}}}    

\begin{table}[htbp]
\centering
\caption{\hl{Overview of Raw Metrics Used by Each Research Question, Clustered by Type. The same color within a cluster reflects correlation.}}
\label{tab:raw-rq-metrics-clustered}

\footnotesize

\setlength{\tabcolsep}{3pt} 

\renewcommand{\arraystretch}{1.1} 

%
%
\begin{tabularx}{\columnwidth}{ p{0.8cm} | p{2.0cm} | >{\raggedright\arraybackslash}X }
\hline
\textbf{RQ} & \textbf{Metric Cluster} & \textbf{Raw Metrics Used} \\ \hline
%
\multirow{2}{*}{\textbf{RQ1}}
& Normalization and Project Size & \hlpurple{Total Lines of Code}, \hlpurple{Non-Comment Line Count}, \hlpurple{Total Character Count}, \hlpurple{Code Character Count}, \hlpurple{Repository Size (MB)} \\ \cline{2-3}
& Descriptive Stats & \hlcyan{Fork Count}, \hlcyan{Star Count}, \hlcyan{Subscriber Count}, \hlorange{Closed PR Count}, \hlorange{Closed Issue Count}, \hlgreen{Open Issue Count}, \hlgreen{Open PR Count}, \hlgreen{Merged PR Count}, \hlred{Comment Line Count}, \hlred{Comment Character Count}, Commit Count, Contributor Count, Core Contributor Count, Average Commit Message Length, README Character Count, Project Start Date, Date of Last Contribution \\ \hline
%
\textbf{RQ2}
& Contribution Activity & Commit Count, \hlblue{Commit Additions}, \hlblue{Commit Deletions} \\ \hline
%
\multirow{3}{*}{\textbf{RQ3}}
& Normalization & \hlpurple{Code Character Count} \\ \cline{2-3}
& Issue Metrics & Qodana Scan Results (Critical, High, Moderate, Low Issues) \\ \cline{2-3}
& Structural Metrics & Understand SciTools Code Metrics (including \hlmagenta{Cyclomatic Complexity}, \hlmagenta{Max Nesting Depth}, Max Inheritance Depth, Comment/Code Ratio, Total Methods/Class, Functions per File, Methods in Class) \\ \hline
\end{tabularx}
\end{table}
\subsubsection{\textbf{Project and Community Characteristics}}
To compare OSS and OSS4SG projects, we analyze 23 metrics capturing distinct aspects of project and community characteristics (Table \ref{tab:project_metrics}). We calculate each metric for every project in our dataset. Since projects vary widely in size, we normalize these metrics by \textit{code\_characters}, the count of source code characters for each project excluding comments, to ensure fair comparison across projects.
For statistical analysis, we use the Mann-Whitney U test to compare metrics between OSS4SG projects and conventional OSS. 
To account for multiple comparisons across 23 metrics, we apply Bonferroni correction, setting our significance threshold at $p < 0.0022$. We complement significance testing with Cliff's $\delta$~\cite{macbeth2011cliff} to measure effect sizes: negligible ($|\delta| < 0.147$), small ($0.147 \leq |\delta| < 0.33$), medium ($0.33 \leq |\delta| < 0.474$), and large ($|\delta| \geq 0.474$). To quantify differences, we compute (OSS4SG/OSS - 1) for each metric. Positive values indicate OSS4SG exceeds conventional OSS, negative values indicate the opposite.


\subsubsection{\textbf{Contributor Overlap}}
\label{subsub:Contributor overlap}

\hl{A common challenge in OSS mining is that developers can commit under multiple identities. To address this, we evaluate three established identity resolution methods. First, we apply email-based matching, which merges contributors sharing identical emails. Second, we use username and email normalization following \citet{zhu2019empirical} which creates identifier pairs by combining normalized emails and usernames. Third, we tested a machine learning approach following ~\citet{amreen2020alfaa}, which measures pair similarity using usernames, emails, and commit fingerprints (commit patterns, file modifications, and activity timestamps). All three methods achieve statistically comparable performance in our dataset (Kruskal-Wallis test \cite{spearman1961general}, p = 0.94), reducing duplicate instances by 9-11\%. We adopt the username and email normalization approach because it correctly identifies true duplicates, such as contributors using both institutional and personal email domains, and maintains deterministic and interpretable behavior.}

Projects within the same ecosystem often attract similar contributors due to shared interests and values. We examine these patterns by analyzing \textit{boundary-spanning contributors}, those who contribute to multiple projects, as they often transfer knowledge and practices across projects~\cite{Crowston2005}.
\hl{A contributor is considered active in multiple projects if they have made at least one commit to two or more projects. To assess the robustness of this definition, we perform a sensitivity analysis using alternative thresholds of 1, 5, and 10 commits.} We measure contributor overlap through two dimensions:
\begin{enumerate}
    \item \textbf{\textsc{Intra-ecosystem overlap:}} The percentage of contributors active in multiple projects within the same ecosystem (either OSS4SG or conventional OSS), indicating how tightly connected each ecosystem is internally.
    \item \textbf{\textsc{Inter-ecosystem overlap:}} The percentage of contributors active in both  OSS4SG projects and conventional OSS, revealing the degree of crossover between mission-driven and conventional development.
\end{enumerate}

We use the Mann-Whitney U test to determine if these overlap patterns differ significantly between OSS and OSS4SG ecosystems.



\subsubsection{\textbf{Contributor Retention And Project Stability:}}

The ability to retain contributors is critical to project stability and long-term survival of any open-source project. However, contributor engagement and retention are not static, they evolve as a project matures~\cite{noei2023empirical, Wang2012}. To provide actionable insights for project maintainers, our analysis aims to uncover how retention patterns differ between OSS4SG and conventional OSS projects across their distinct life stages. This approach allows us to identify precisely when these two types of projects diverge and at which stages retention efforts are most crucial.

\paragraph{Project Stability}
To assess a project's stability profile, we adopt the quadrant model from ~\citet{Ferreira2020}, which relies on quarterly contributor turnover.
We use 3-month windows for this analysis, a time frame consistent with prior work on OSS group dynamics~\cite{bock2021measuring}. For each 3-month window  $t$, we compute the Join Rate and Leave Rate:

\[
\text{Join Rate}(t) = \frac{\text{\# New Contributors in Window } t}{\text{Total Contributors Active in Window } t}
\]
\[
\text{Leave Rate}(t) = \frac{\text{\# Contributors Leaving in Window } t}{\text{Total Contributors Active in Window } t}
\]

\vspace{2mm}

A contributor is considered to have left if they have not committed for at least five months after window $t$, distinguishing true departures from brief breaks~\cite{jamieson2024predicting}. 

We then classify a project's stability by using the median join and leave rates as thresholds per category. This yields four profiles: \textsc{Stable} (low join, low leave), \textsc{Unstable} (high join, high leave), \textsc{Attractive} (high join, low leave), and \textsc{Unattractive} (low join, high leave). From these constructs, we define \emph{magnetic} projects as those with high join rates (Unstable and Attractive) and \emph{sticky} projects as those with low leave rates (Stable and Attractive). Finally, we use Mann-Whitney U tests to assess whether the join and leave rates differ significantly between OSS4SG and conventional OSS projects.

\paragraph{Contributor Retention Over Time}
To analyze how retention evolves over time, we follow a four-step methodology. First, We segment the timeline of each project into \emph{early}, \emph{middle}, and \emph{late} stages based on its age quartiles, following established life stage models~\cite{noei2023empirical, Wang2012}. 
\hl{To ensure these stages are comparable across projects, we only include repositories with a lifespan of at least seven years, the first-quartile age for our dataset. }
. Second, we measure retention within each stage using a \textit{Retention Ratio}, calculated over three-month rolling windows ~\cite{bock2021measuring}. The ratio is calculated as:
\[
\text{Retention Ratio}(i) = \frac{\text{Join Rate}(i) - \text{Leave Rate}(i)}{\text{Join Rate}(i) + \text{Leave Rate}(i)}
\]
This metric ranges from -1 (indicating more contributors left than joined) to +1 (indicating more contributors joined than left), with values near zero suggesting a balanced turnover. Third, we group the calculated retention ratios into six distinct categories (OSS/OSS4SG $\times$ early/mid/late). Finally, we apply the Scott-Knott test~\cite{scott1974cluster, noei2023empirical} to partition these six groups into statistically distinct clusters, revealing significant differences in retention performance and rank the retention ratio among six groups. \hl{The Scott-Knott-ESD test incorporates both statistical significance and effect size, separating groups only when differences exhibit medium or large effect sizes \cite{pinto2016more}, which reflects realistic dynamics in practice \cite{israel1992determining,zhang2014towards}}

\subsection{Findings}
\label{rq1:findings}

\begin{table}[htbp]
\centering
\caption{\hl{Mann--Whitney U Test Results for Project and Community Metrics, with correlation cluster ID. We show statistically significant differences (green cells), effect size (L/M/S), and (OSS4SG/OSS -1).}}
\label{tab:project_metrics}
\resizebox{\columnwidth}{!}{%
\begin{tabular}{c|l|c|c|c}
\hline
\textbf{ID} & \textbf{Metric} & \textbf{p-value} & \textbf{Effect Size} & \textbf{OSS4SG/OSS Ratio} \\
\hline

\rowcolor{gray!20}\multicolumn{5}{l}{\textbf{Popularity}} \\ \hline
\cellcolor{blue!20}1 & Star Count & \cellcolor{green!20}0.00 & L & -0.80 \\ \hline
\cellcolor{blue!20}1 & Subscriber Count & \cellcolor{green!20}0.00 & L & -0.70 \\ \hline
\cellcolor{blue!20}1 & Fork Count & \cellcolor{green!20}0.00 & L & -0.76 \\ \hline

\rowcolor{gray!20}\multicolumn
{5}{l}{\textbf{Community Composition}} \\ \hline
0 & Contributor Count & 0.58 & S & -0.45 \\ \hline
0 & Core Contributor Count & \cellcolor{green!20}0.00 & M & 1.80 \\ \hline

\rowcolor{gray!20}\multicolumn{5}{l}{\textbf{Contributions}} \\ \hline
0 & Commit Count & \cellcolor{green!20}0.00 & M & 0.90 \\ \hline
\cellcolor{green!30}3 & Open PR Count & 0.15 & S & -0.20 \\ \hline
\cellcolor{orange!20}2 & Closed PR Count & \cellcolor{green!20}0.00 & M & 2.80 \\ \hline
0 & Merged PR Count & 0.15 & S & 0.15 \\ \hline
\cellcolor{green!30}3 & Open Issue Count & 0.11 & S & 0.10 \\ \hline
\cellcolor{orange!20}2 & Closed Issue Count & \cellcolor{green!20}0.00 & S & 2.00 \\ \hline
0 & Average PR Resolution Time & \cellcolor{green!20}0.00 & S & -0.31 \\ \hline

\rowcolor{gray!20}\multicolumn{5}{l}{\textbf{Readability}} \\ \hline
0 & Average Commit Message Length & \cellcolor{green!20}0.00 & S & 0.50 \\ \hline
\cellcolor{pink!20}5 & Comment Line Count & 0.98 & S & -0.42 \\ \hline
\cellcolor{pink!20}5 & Comment Character Count & 0.57 & S & 0.92 \\ \hline
0 & Code-to-Comment Ratio & 0.07 & S & 0.78 \\ \hline
0 & Code-to-Comment Character Ratio & 0.60 & S & -0.30 \\ \hline

\rowcolor{gray!20}\multicolumn{5}{l}{\textbf{Project Size}} \\ \hline
\cellcolor{red!20}4 & Total Lines of Code & 0.34 & S & -0.01 \\ \hline
\cellcolor{red!20}4 & Non-Comment Line Count & 0.31 & S & 0.16 \\ \hline
\cellcolor{red!20}4 & Total Character Count & 0.19 & S & 0.18 \\ \hline
\cellcolor{red!20}4 & Code Character Count & 0.17 & S & 0.31 \\ \hline
0 & \textsc{Readme} Character Count & 0.29 & S & -0.35 \\ \hline
\cellcolor{red!20}4 & Repository Size (MB) & \cellcolor{green!20}0.00 & S & -0.08 \\ \hline

\end{tabular}%
}
\end{table}

\subsubsection{\textbf{Project Characteristics}}
\label{RQ1:project characteristics}
\takeaway{OSS4SG and OSS differ significantly in popularity, community composition, contributions, review process, readability, and project size.} Table~\ref{tab:project_metrics} shows the results of our comparison.

\hl{We identify five highly correlated metric clusters, as shown by the colour-coded IDs in Table\ref{tab:project_metrics}. Popularity metrics (i.e., Star Count, Subscriber Count, and Fork Count) are all strongly correlated. In all cases OSS projects show higher popularity values than OSS4SG projects, so retaining one representative metric such as Star Count is sufficient as a proxy to explain popularity differences. Within contribution metrics, Open PR Count and Open Issue Count are highly correlated, but neither show a significant difference between OSS and OSS4SG ecosystems. Similarly, Closed Issue Count and Closed PR Count show high correlations; therefore, one of them can be retained as representative of the cluster. Readability metrics such as Comment Line Count and Comment Character Count are correlated but reveal no significant differences across project types. Finally, project size metrics including Total Lines of Code, Non-Comment Line Count, Total Character Count, Code Character Count, and Repository Size are highly correlated, showing consistent patterns without statistically meaningful differences. Therefore, Repository Size or Readme Character Count can serve as a single proxy for project size.}

\textsc{\textbf{Popularity:}} We use repositories stars, forks and subscriber count as proxy for popularity (see Table \ref{tab:project_metrics}). 
Our findings show that conventional \takeaway{OSS projects are significantly more popular than OSS4SG projects across all popularity metrics with a large effect size}, indicating that the difference is not only statistically significant but also practically meaningful in terms of visibility and external interest. This observation aligns with prior work showing that OSS4SG, despite its societal importance, generally receives less public exposure~\cite{Huang2021}.

\textsc{\textbf{Community composition:}} \hl{Consistent with prior work \cite{Huang2021}, the total number of contributors normalized shows no statistically significant difference between OSS and OSS4SG projects. However, OSS4SG projects have a significantly larger core contributor base, 1.8 times higher than conventional OSS.}
This suggests that OSS4SG communities may foster deeper, more sustained engagement, likely driven by contributors’ intrinsic motivations and strong personal commitment to social impact over technical recognition or external incentives.

\textsc{\textbf{Contributions:}} \takeaway{OSS4SG projects exhibit significantly higher commit counts—twice as many as conventional OSS (Mann– Whitney U, p < 0.001,  medium effect size), despite having fewer total contributors}  (see Table \ref{tab:project_metrics}). This likely results from the higher proportion of core contributors in OSS4SG communities, which drives more consistent and sustained development activity.

\textsc{\textbf{Review process:}} Table~\ref{tab:project_metrics} reveals that \takeaway{OSS4SG projects close significantly more pull requests and resolve them faster, indicating more efficient review workflows.} Although not statistically significant, OSS4SG projects also demonstrate a 15\% higher pull request acceptance rate compared to conventional OSS projects (see Table~\ref{tab:project_metrics}). This observation aligns with prior literature reporting higher acceptance rates for student-submitted pull requests within OSS4SG communities~\cite{fang2023four}, suggesting a potentially more inclusive and responsive review process.


\textsc{\textbf{Readability:}} \takeaway{OSS4SG projects have a higher comment line and comment character count.} Out of the various readability metrics, the average commit message length is statistically significant (Mann–Whitney U, $p < 0.001$, small effect size). Indeed, commit message in OSS4SG projects are 50\% longer compared to commit messages in conventional OSS projects. This result suggests that OSS4SG contributors might prioritize transparent communication and a more inclusive collaboration.

\subsubsection{\textbf{Contributor Overlap}}
\takeaway{OSS4SG projects have a smaller but more interconnected contributor community compared to conventional OSS projects.} Although conventional OSS project have 45\% more contributors per projects, OSS4SG projects exhibit significantly higher \textsc{Intra-ecosystem overlap} (see section \ref{subsub:Contributor overlap}).

In fact,\hl{ 10\%} of OSS4SG contributors actively participate in two or more OSS4SG projects, compared to only \hl{3\%} within conventional OSS (Mann–Whitney U, $p < 0.0001$, small effect size).

\hl{Our threshold sensitivity analysis confirms that OSS4SG exhibits higher intra-ecosystem overlap across different commit thresholds. Specifically, OSS4SG projects exhibit 3.33 times higher contributor interconnectedness when considering contributors with at least one commit, 3.40 times higher at five commits, and 6.63 times higher at ten commits. This trend indicates that as participation intensity increases, OSS4SG contributors form even tighter collaboration networks.}

\textsc{Inter-ecosystem overlap}, however, displays similar rates for conventional OSS \hl{(2.4\%)} and OSS4SG \hl{(2.63\%)}. This may be attributed to the inherent social nature of OSS4SG projects, which tend to attract contributors primarily motivated by societal benefit \cite{Huang2021}. In contrast, traditional OSS projects often draw individuals who contribute to OSS for their own use in a ``scratch your own itch'' fashion \cite{gerosa2021shifting}  \cite{roberts2006understanding} or seek to develop specific technical skills \cite{Huang2021, gerosa2021shifting}.

\begin{tcolorbox}[
    colback=gray!10,
    colframe=gray!10,
    colbacktitle=gray!60, 
    coltitle=white, 
    title=RQ1 Summary: Project Characteristics,
    sharp corners,
    toptitle=1ex  
]

Despite conventional OSS projects being significantly more popular, OSS4SG projects have 1.8x more core contributors and 1.9x higher commit frequency.
Intra-ecosystem collaboration is significantly more prominent in OSS4SG projects with 10.23\% contributors actively participating in multiple OSS4SG projects compared to only 3.07\% in conventional OSS.
\end{tcolorbox}

\begin{figure}[htbp]
\centering
\includegraphics[width=4cm]{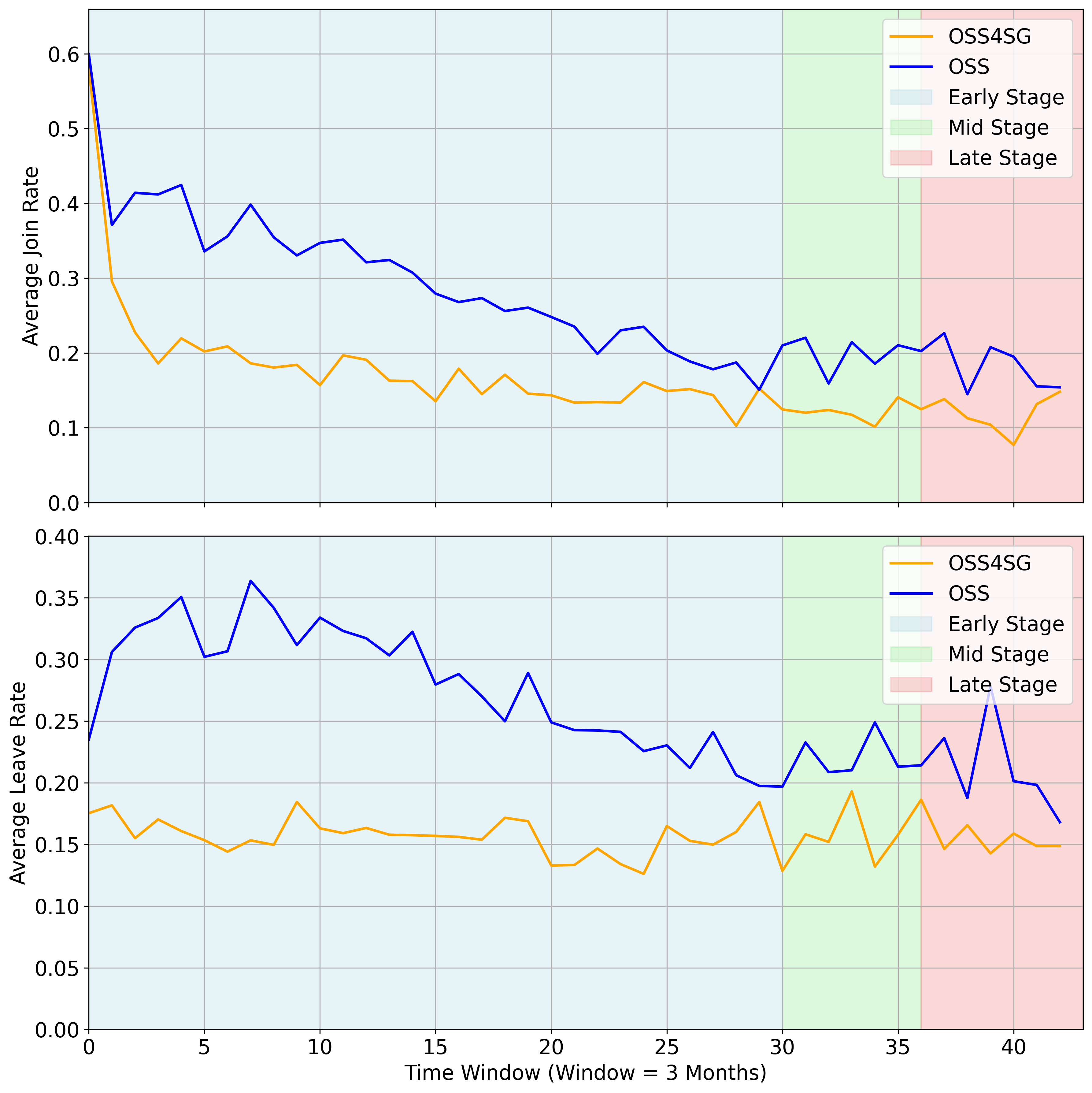}
\caption{Time-series comparison of join and leave rates for OSS4SG and conventional OSS  projects. The background color indicates early, mid, and late stages. }
\label{fig:RQ2 join and leave rate through project life stages}
\end{figure}

\begin{figure}[!htbp]
    \centering
    \includegraphics[width=7.5cm]{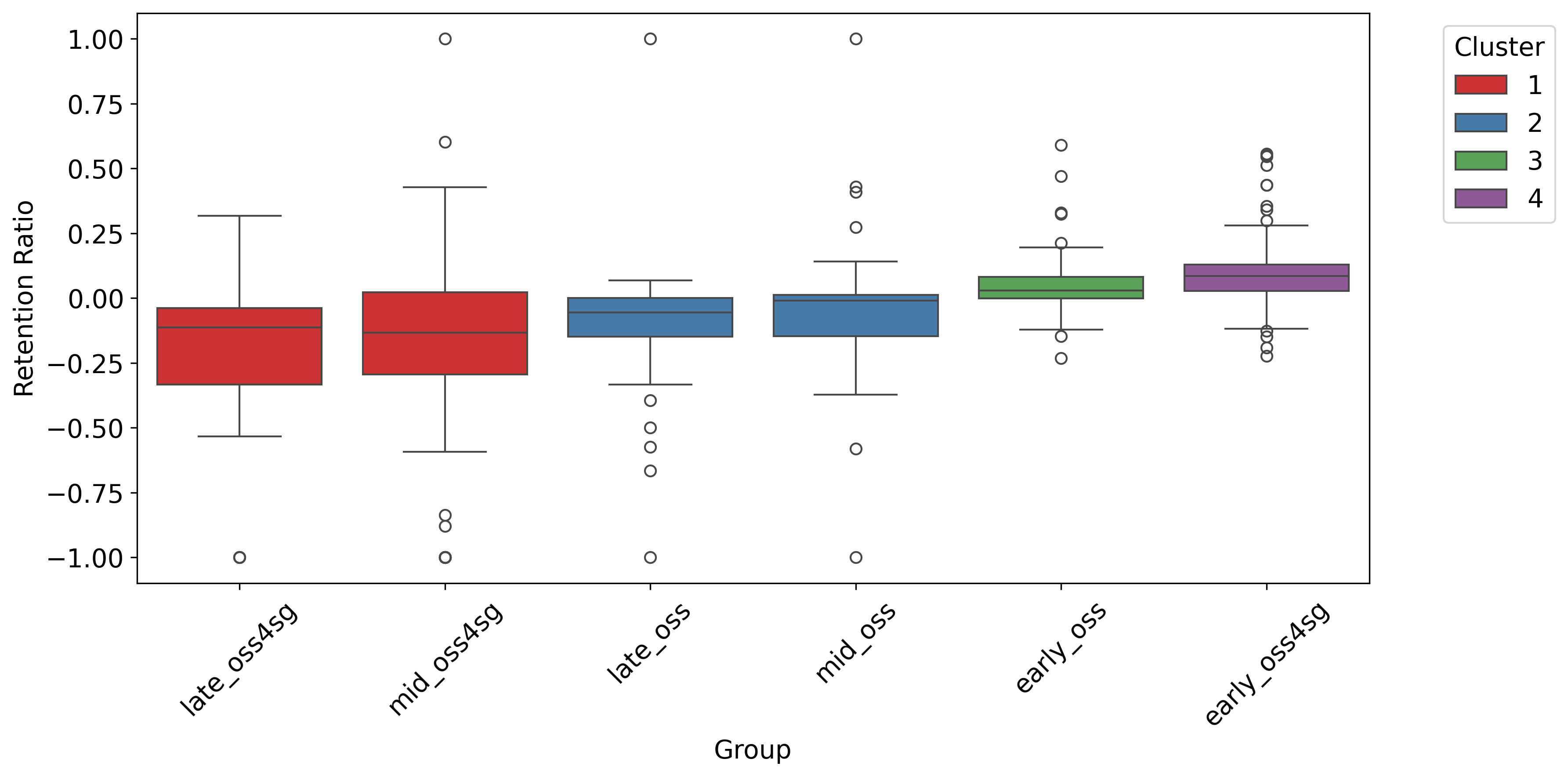}
     \caption{Scott-Knott clusters Comparison of retention ratios between OSS4SG and conventional OSS projects.}
    \label{fig:core_vs_noncore_boxplot_scott_knott}
\end{figure}

\begin{figure}[!htbp]
\centering
\includegraphics[width=0.9\linewidth]{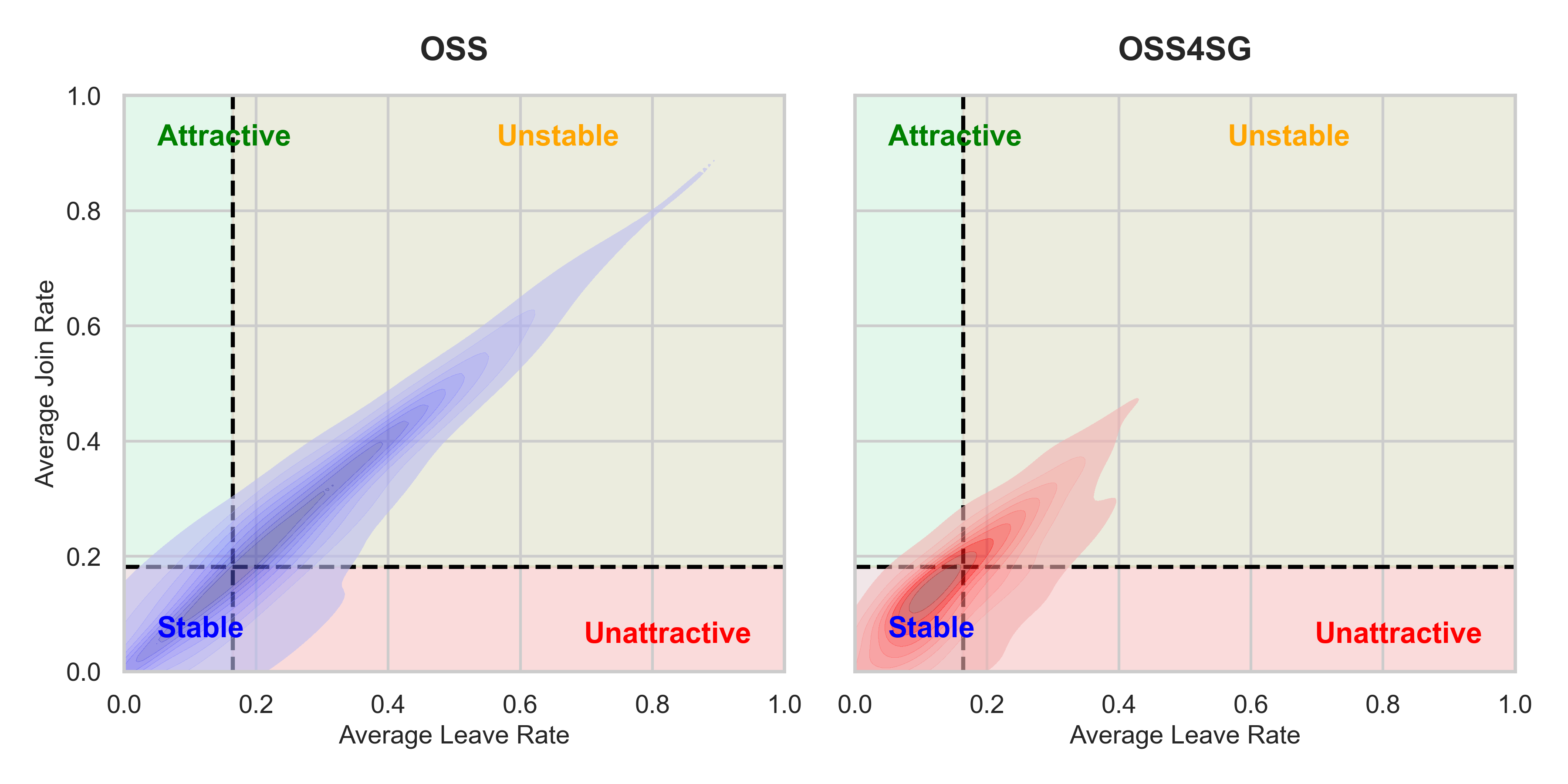}
\caption{Visualization of the four quadrants classification using median splits of the join and leave rates in OSS4SG and OSS projects.
\hl{Kernel density estimation (KDE) \cite{wiki:KernelDensityEstimation} is used to represent project distributions, where darker regions indicate higher project density.}}
\label{fig:RQ1 KDE plot for projects stability}
\end{figure}

\subsubsection{\textbf{Contributors Retention And Project Stability}}
\label{RQ1:retention and stability}

\paragraph{\textbf{Retention:}} \takeaway{Conventional OSS projects have high early-stage join rates but also higher turnover that progressively steadies. In contrast, OSS4SG projects exhibit relatively consistent join and leave rates throughout all life stages.} Figure~\ref{fig:RQ2 join and leave rate through project life stages} shows the overall trend of the join and leave rates throughout the OSS4SG and conventional OSS projects life stages. The higher early-stage join rate in conventional OSS projects may be attributed to their significantly greater popularity (see Table \ref{tab:project_metrics}). However, this influx is accompanied by a comparably high leave rate, indicating a more volatile engagement pattern and challenges in retaining contributors beyond initial interest. In contrast, OSS4SG projects maintain steadier join and leave rates, even in early stages, suggesting deeper personal investment and more consistent commitment from contributors (see Figure \ref{fig:RQ2 join and leave rate through project life stages}).

The Scott-Knott statistical clustering further confirms the above difference between OSS and OSS4SG throughout the life stages of projects (see Figure ~\ref{fig:core_vs_noncore_boxplot_scott_knott}). Specifically, OSS4SG projects in their early stage demonstrate the highest retention ratios (cluster 4), significantly outperforming early-stage conventional OSS projects (cluster 3). \hl{For instance, when comparing early-stage projects, OSS4SG projects exhibit retention ratios of 8\% compared to 3\% in conventional OSS, a statistically significant difference with medium-to-large effect size. This pattern is further illustrated by examples such as oppia/oppia \cite{OppiaTeam2013Oppia} OSS4SG project, where 25\% of developers stop contributing, compared to 62\% in the philc/vimium \cite{Crosby2010Vimium} SS project, demonstrating the higher turnover characteristic of conventional OSS.} As projects mature (i.e., mid and late stages), both conventional OSS and OSS4SG projects exhibit reduced contributor retention, with similar patterns emerging across clusters 1 and 2. 
\paragraph{\textbf{Stability:}} \takeaway{OSS4SG projects maintain significantly more stable contributor dynamics compared to conventional OSS projects, with strong statistical significance ($p < 0.001$) and a large effect size}. As shown in Figure~\ref{fig:RQ1 KDE plot for projects stability}, the majority of OSS4SG projects (60\%) fall into the \emph{Stable} quadrant (low join, low leave). In contrast, most OSS projects (74\%) are classified as \emph{Unstable} (high join, high leave), suggesting higher contributor turnover. 
\takeaway{Overall, 63.4\% of OSS4SG projects form \emph{sticky} communities (stable and attractive), whereas 75.4\% of conventional OSS projects are \emph{magnetic} (attractive or unstable). }
These findings highlight that OSS4SG projects tend to keep their existing contributors more effectively, while conventional OSS often compensates for high turnover by attracting new members. The stronger stability observed in OSS4SG projects primarily stems from early-stage contributor retention, highlighting the importance of the first few years in a project's life and the continuous engagement enabled by social-good missions.




\begin{tcolorbox}[
    colback=gray!10,
    colframe=gray!10,
    colbacktitle=gray!60, 
    coltitle=white, 
    title=RQ1 Summary: Retention \& stability,
    sharp corners,
    toptitle=1ex  
]
OSS4SG projects form sticky communities (63.4\% stable and attractive) while conventional OSS projects are predominantly magnetic (75.4\% attractive or unstable).
OSS4SG shows higher retention, particularly in early stages, with more consistent engagement compared to conventional OSS.
\end{tcolorbox}



\section{Patterns and Evolution of Contributor Engagement (RQ2)}
\label{subsec:contributor engagement }

\subsection{Motivation}
Our analysis in RQ1 (see Section \ref{rq1:findings}) revealed that OSS and OSS4SG projects have different community structures and stability profiles from the outside. This raises a crucial question about the people on the inside: does a contributor drawn to a social-good mission behave differently than one motivated by technical challenges? To find out, we now shift our focus from the project-level to the contributor-level. This section investigates the internal development dynamics by analyzing the engagement patterns of different contributor types to determine whether the mission-driven nature of OSS4SG fosters a more consistent and dedicated contributor.

\subsection{Approach}
To answer this, we conduct three distinct analyses. We:
(1) compare daily commit frequencies to measure temporal engagement patterns, 
(2) calculate a core development ratio to quantify how much development work core contributors handle, and 
(3) use time-series clustering to identify distinct engagement patterns across project life stages.

\subsubsection{\textbf{Temporal Engagement and Core Development Ratio Analysis}}

To analyze the annual contribution patterns, we first identified the core and casual contributors for each individual project in our dataset. Following the 80/20 rule commonly used in prior work~\cite{Ferreira2020}, we classified the top 20\% of a project's contributors who were responsible for 80\% of its total commits as \textit{core contributors}. All other contributors for that project were labelled as \textit{casual}. To quantify the common contribution engagement patterns, we measure core contributor engagements by analyzing both commit activity and code churn (sum of insertions and deletions)\cite{noei2023empirical, olsson2017relationship}.\hl{Although commit additions and deletions are highly correlated, both are necessary to calculate code churn, which captures the complete scope of development effort.} To this end, for each project, we compute code churn by core and casual contributors for each window. Then define:

\[
\text{Core Development Ratio(i)} 
= \frac{\text{Core churn (i)}}{\text{Core churn(i)+} \text{Casual churn(i)}}
\]

where \(\text{Core churn}(i)\) and \(\text{casual churn}(i)\) represent the total code churn (lines added and deleted) from core and casual contributors, respectively, within a given project \(i\). The resulting ratio indicates the proportion of development activity driven by the project's core members. With each contributor labeled, we aggregated all commits from all projects. For each of the 365 days of the year, we calculated the total number of commits made across all projects, grouping them into four distinct categories: core contributors in OSS4SG, casual contributors in OSS4SG, core contributors in conventional OSS, and casual contributors in conventional OSS. To enable a direct comparison of engagement trends between these groups, which have different total commit volumes, we normalized each of the four daily time-series independently using min-max scaling. This process converts the daily commit counts for each group to a scale of 0 to 1, allowing for a clear visual comparison of their temporal engagement patterns.
We also examine weekend contribution patterns (Saturday and Sunday) to assess engagement beyond typical working days.


\subsubsection{\textbf{Time-Series Clustering Of Engagement Patterns}}

We segment each project’s time series (of \textit{CoreDevRatio}) by developmental stage and apply time-series clustering using \textit{Soft Dynamic Time Warping (Soft-DTW)} as the distance measure. Unlike classical DTW, which finds an optimal alignment path through hard constraints, Soft-DTW introduces a differentiable and more stable formulation that better handles partially misaligned temporal sequences. This makes it particularly suitable for capturing nuanced similarities in contributor engagement across projects with varying lifespans and activity rhythms. By clustering in Soft-DTW space, we group projects based on shared temporal \textit{patterns} of core activity.

\subsection{Findings}

\begin{figure}[!htbp]
\centering
\includegraphics[width=5cm]{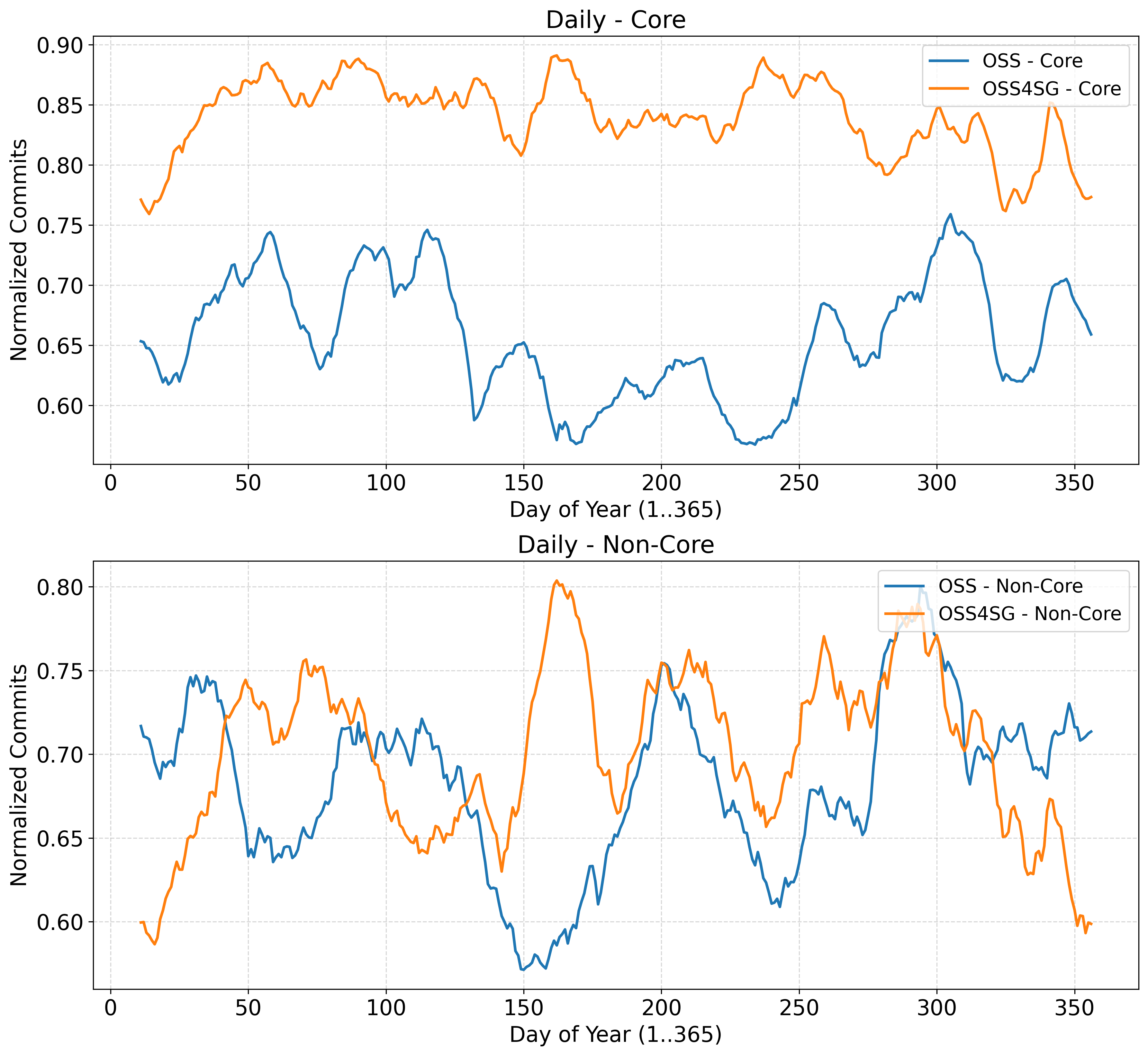}
\caption{Commit frequency plots showing time-series comparisons between OSS4SG and conventional OSS for core and casual contributors.}
\label{fig:RQ2 commit frequency core vs. non-core}
\end{figure}

\subsubsection{\textbf{Commit Frequency And Core Contributors Activity}}
\label{RQ2: Commit Frequency and core contributors activity}
When investigating a project's contributor base, engagement is equally important to retention. In fact, higher levels of consistent contribution are critical to project's health and sustainability \cite{xiao2023early}.

\begin{figure}[!!htbp]
    \centering
    \includegraphics[width=6cm]{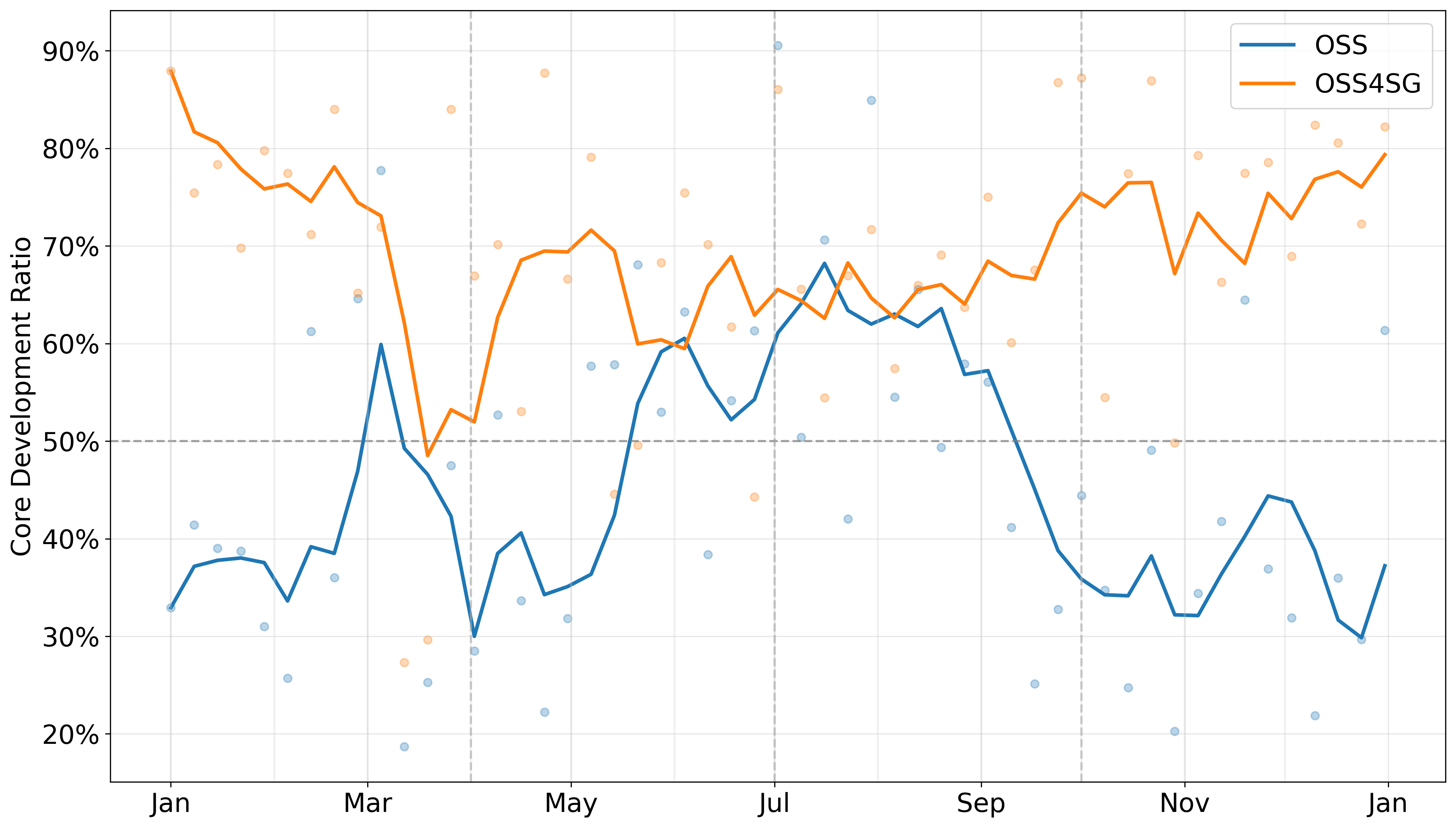}
    \caption{Annual pattern of core developer contribution ratio (code churn).}
    \label{fig:RQ2 core code churn}
\end{figure}

\begin{figure}[!htbp]
    \centering
    \includegraphics[width=0.8 \columnwidth]{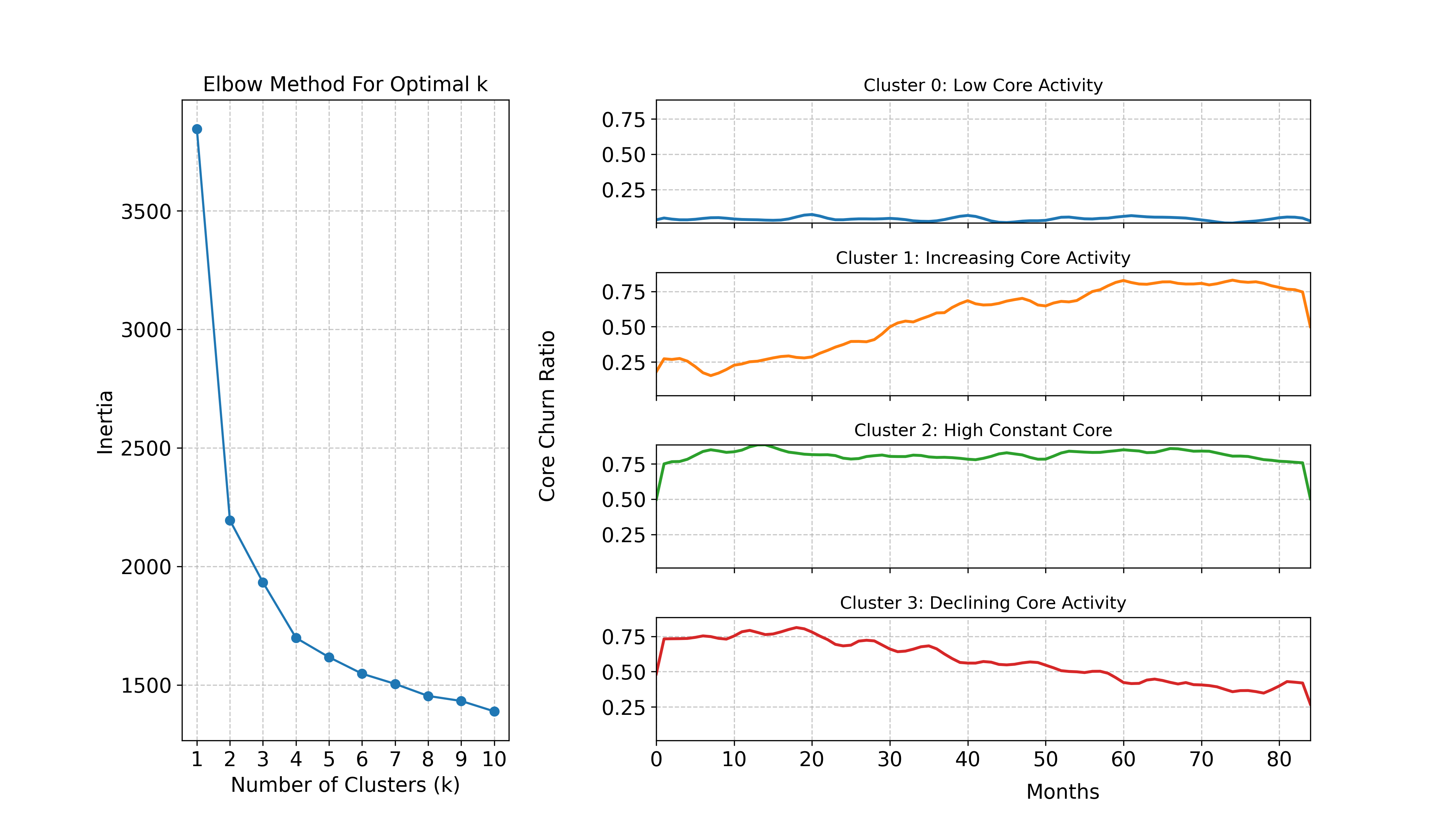}
    \caption{Clusters of core contributor behavior in OSS and OSS4SG projects and the elbow curve.}
    \label{fig:core code churn ration clusters}
\end{figure}

Figure~\ref{fig:RQ2 commit frequency core vs. non-core} compares normalized daily commit frequencies for OSS4SG vs.\ conventional OSS. We find that \takeaway{OSS4SG core contributors exhibit significantly higher and more consistent commit activity (Mann--Whitney U, p < 0.001, medium effect size)}, often maintaining above 80\% of project commits year-round with minimal seasonal dips. By contrast, OSS core contributors show greater fluctuations, at times dropping below 60\% of project commits. This difference suggests a more sustained commitment from core contributors.

\begin{table}[htbp]
\centering
\footnotesize
\setlength{\tabcolsep}{3pt}
\renewcommand{\arraystretch}{1.0}
\caption{Core Contributor Activity Patterns disaggregated by Project Stage.}
\label{tab:cluster_percentages_single_line}
\begin{tabular}{lcccccc}
\toprule
\textbf{Cluster} & \multicolumn{2}{c}{\textbf{Early}} & \multicolumn{2}{c}{\textbf{Mid}} & \multicolumn{2}{c}{\textbf{Late}} \\
\cmidrule(l){2-3} \cmidrule(l){4-5} \cmidrule(l){6-7}
 & OSS & OSS4SG & OSS & OSS4SG & OSS & OSS4SG \\
\midrule
Low Core Activity        & 26\% & 10\% & 34\% & 27\% & 17\% & 20\% \\
Increasing Core Activity & 18\% & 18\% &  6\% & 11\% &  8\% & 25\% \\
High Constant Core       & \textbf{33\%} & \textbf{45\%} & \textbf{41\%} & \textbf{37\%} & 25\% & \textbf{55\%} \\
Declining Core Activity  & 23\% & 27\% & 19\% & 25\% & \textbf{50\%} &  0\% \\
\bottomrule
\end{tabular}
\normalsize
\end{table}

The normalized weekend contribution patterns further highlight this distinction. 
\takeaway{OSS4SG core contributors contribute 35\% more commits during weekends (Saturdays and Sundays) compared to conventional OSS core contributors}. This difference is both statistically significant and demonstrates a large effect size, indicating meaningful practical impact (Mann–Whitney U, $p < 0.00001$, Large effect size).This engagement might stems from a deeper connection to the project's mission, motivating OSS4SG contributors to invest their personal time.

 Casual contributor activity exhibits comparable seasonality in both OSS4SG and conventional OSS (Mann--Whitney U, $p=0.0084$, medium effect size), with peaks in June and October mapping to major events like Google Summer of Code and Hacktoberfest.Many projects in OSS4SG and conventional OSS have GitHub topics related to Hacktoberfest, which led us to investigate these events. 
 We find that, new OSS4SG contributors who join during these events maintain a higher post-event commit rate (0.09 commits/day) compared to OSS newcomers (0.04 commits/day). This difference suggests that social-good missions may foster continued engagement beyond the event period.

To further confirm these observations, Figure~\ref{fig:RQ2 core code churn} illustrates weekly aggregated code churn attributed to core contributors. \takeaway{OSS4SG projects consistently maintain a core code churn above 50\% compared to conventional OSS, falling under 50\% code churn most of the year around.} Collectively, these findings underscore the distinct engagement patterns of OSS4SG core contributors, characterized by higher consistency, lower seasonality, and enhanced weekend activity, compared to conventional OSS projects.

\subsubsection{\textbf{Time-Series Clustering Of Core Development Ratios}}
\label{RQ2:Time series clustring}
Our clustering analysis of core development ratios reveals four distinct patterns in how projects split coding efforts between core and casual contributors (see Figure~\ref{fig:core code churn ration clusters}). Our clustering achieved a Silhouette score of 0.404  with K = 4, which measures how well our data points fit into their assigned clusters. This score ranges from -1 (poor fit) to 1 (excellent fit). Table \ref{tab:cluster_percentages_single_line} shows that in \textit{High Constant Core (Cluster~2)}, core contributors consistently drive the bulk of code changes across the entire project lifespan. 
\takeaway{The High Constant Core pattern is more prevalent in OSS4SG}, where 45\% of early-stage projects, 37\% of mid-stage projects, and 55\% of late-stage projects fall into this cluster, compared to only 33\%, 41\%, and 25\% in conventional OSS.

By contrast, \takeaway{OSS projects show  instability in later stages, with 50\% of projects displaying a \textit{Declining Core Activity pattern (Cluster~3) }, compared to 0\% of OSS4SG projects}. This difference suggests that mission-driven project may better retain core expertise as they mature. Furthermore, even in early stages, 45\% of OSS4SG projects already exhibit strong \textit{High Constant Core} engagement, whereas 27\% of OSS projects instead remain in \textit{Low Core Activity (Cluster~0)}, double the 11\% in OSS4SG. By late stages, more than half (55\%) of OSS4SG projects sustain high core engagement (Cluster~2), compared to 25\% of OSS. These trajectories build upon RQ1's (see Section \ref{RQ1:retention and stability}) stability findings: OSS4SG's ``sticky''communities map to persistent core leadership, while OSS's ``magnetic-but-unstable'' map to declining core activity.


\begin{tcolorbox}[
    colback=gray!10,
    colframe=gray!10,
    colbacktitle=gray!60, 
    coltitle=white, 
    title=RQ2 Summary,
    sharp corners,
    toptitle=1ex  
]
OSS4SG projects consistently maintain High Constant Core patterns (55\% in late stages vs 25\% in OSS), with core contributors sustaining 80\%+ engagement year-round and 35\% higher weekend activity compared to conventional OSS.
As projects mature (i.e late stage), 50\% of conventional OSS projects have the Declining Core Activity pattern compared to 0\% in OSS4SG.

\end{tcolorbox}
    

\newcommand{\rankA}[1]{\cellcolor{Green!25}{#1}}       
\newcommand{\rankB}[1]{\cellcolor{GreenYellow!25}{#1}} 
\newcommand{\rankC}[1]{\cellcolor{Yellow!25}{#1}}      
\newcommand{\rankD}[1]{\cellcolor{Orange!25}{#1}}      
\newcommand{\rankE}[1]{\cellcolor{Red!25}{#1}}         
\newcommand{\rankF}[1]{\cellcolor{Red!40}{#1}}         
\newcommand{\R}[2]{
  \ifcase#1 \or\rankA{#2}\or\rankB{#2}\or\rankC{#2}\or
            \rankD{#2}\or\rankE{#2}\else\rankF{#2}\fi}
\begin{table*}
\centering
\caption{The results of the Scott–Knott test on our metrics for each core contribution pattern. Same colors indicate statistically indistinguishable clusters. Abbreviations: I = increasing core activity, D = declining core activity, H = high constant core activity, and L = low core activity.}
\resizebox{\linewidth}{!}{%
\begin{tabular}{llcccccccc}
\toprule
\textbf{Category} & \textbf{Metric} &
\multicolumn{8}{c}{\textbf{Same color = Same Rank}}\\
\midrule
\multirow{3}{*}{Complexity}
 & Cyclomatic &
\R{4}{OSS-D (4)} & \R{3}{4SG-L (3)} & \R{2}{4SG-D (2)} & \R{2}{OSS-L (2)} &
\R{2}{4SG-I (2)} & \R{2}{4SG-H (2)} & \R{2}{OSS-I (2)} & \R{1}{OSS-H (1)}\\
 & Max Inheritance &
\R{4}{OSS-I (4)} & \R{4}{OSS-D (4)} & \R{3}{4SG-L (3)} & \R{3}{OSS-H (3)} &
\R{3}{4SG-I (3)} & \R{2}{OSS-L (2)} & \R{1}{4SG-H (1)} & \R{1}{4SG-D (1)}\\
 & Max Nesting &
\R{4}{4SG-H (4)} & \R{4}{4SG-I (4)} & \R{4}{OSS-L (4)} & \R{4}{4SG-D (4)} &
\R{3}{OSS-D (3)} & \R{3}{OSS-H (3)} & \R{2}{4SG-L (2)} & \R{1}{OSS-I (1)}\\
\midrule
\multirow{1}{*}{Documentation}
 & Comment/Code &
\R{1}{OSS-L (1)} & \R{1}{4SG-D (1)} & \R{1}{OSS-I (1)} & \R{2}{OSS-D (2)} &
\R{2}{OSS-H (2)} & \R{3}{4SG-H (3)} & \R{3}{4SG-I (3)} & \R{3}{4SG-L (3)}\\
\midrule
\multirow{3}{*}{Organization}
 & Methods &
\R{1}{4SG-L (1)} & \R{1}{4SG-H (1)} & \R{1}{4SG-I (1)} & \R{1}{OSS-H (1)} &
\R{2}{OSS-D (2)} & \R{2}{OSS-I (2)} & \R{3}{OSS-L (3)} & \R{3}{4SG-D (3)}\\
 & Functions &
\R{1}{4SG-H (1)} & \R{1}{4SG-L (1)} & \R{1}{4SG-I (1)} & \R{2}{OSS-L (2)} &
\R{2}{OSS-I (2)} & \R{3}{4SG-D (3)} & \R{4}{OSS-D (4)} & \R{4}{OSS-H (4)}\\
 & Decl.\ Methods &
\R{1}{4SG-I (1)} & \R{1}{4SG-H (1)} & \R{2}{OSS-H (2)} & \R{2}{4SG-L (2)} &
\R{3}{OSS-D (3)} & \R{4}{4SG-D (4)} & \R{4}{OSS-L (4)} & \R{4}{OSS-I (4)}\\
\midrule
\multirow{3}{*}{Issue Severity}
 & Critical &
\R{5}{4SG-D (5)} & \R{4}{4SG-L (4)} & \R{4}{4SG-I (4)} & \R{4}{OSS-H (4)} &
\R{4}{OSS-I (4)} & \R{3}{4SG-H (3)} & \R{2}{OSS-D (2)} & \R{1}{OSS-L (1)}\\
 & High &
\R{5}{OSS-D (5)} & \R{4}{4SG-L (4)} & \R{3}{OSS-I (3)} & \R{3}{OSS-H (3)} &
\R{2}{4SG-D (2)} & \R{2}{4SG-I (2)} & \R{2}{4SG-H (2)} & \R{1}{OSS-L (1)}\\
 & Moderate &
\R{5}{4SG-L (5)} & \R{5}{4SG-I (5)} & \R{4}{OSS-I (4)} & \R{3}{4SG-D (3)} &
\R{3}{OSS-L (3)} & \R{2}{OSS-D (2)} & \R{2}{OSS-H (2)} & \R{1}{4SG-H (1)}\\
\bottomrule
\end{tabular}}
\label{tab:sk_rank_by_metric}
\end{table*}

\section{Code quality and issue handling (RQ3)}

\subsection{Motivation}

High developer turnover can negatively impact code quality and slow issue resolution times~\cite{foucault2015impact, ferreira2020turnover, nassif2017revisiting}. Our findings from RQ1 (see Section \ref{RQ1:retention and stability}) and RQ2 (see Section \ref{RQ2: Commit Frequency and core contributors activity}) reveal that OSS4SG and conventional OSS projects exhibit different contributor engagement and stability patterns, which could lead to  differences in code quality and issue resolution velocity. Understanding these relationships will help project maintainers optimize their contributor management strategies based on their project's contributor base characteristics.

\subsection{Approach}
\label{sec:rq3_approach}

We measure code quality using static analysis tools and map quality changes to the contribution patterns identified in RQ2 (see Section \ref{RQ2:Time series clustring}).

\subsubsection{\textbf{Code Quality Measurement Tools}}
\label{sec:rq3_tools}

We use two complementary static analysis tools to measure code quality from different perspectives.
We use JetBrains' Qodana~\cite{Qodana} to identify and categorize code quality issues. Qodana scans the codebases and flags problems related to bugs, and security vulnerabilities. It classifies each issue into four severity levels: \emph{critical}, \emph{high}, \emph{moderate}, and \emph{low}. Table~\ref{tab:qodana_severities} defines each severity level with examples.

We employ SciTools' Understand~\cite{Understand} to extract structural metrics from source code. These static metrics quantify architectural and complexity characteristics without code execution. We have selected six widely-used metrics that represent key facets of code complexity, documentation, and organization~\cite{noei2025empirical}. Table~\ref{tab:understand-metrics} lists each selected metric and their definitions.\hl{Although Cyclomatic Complexity and Max Nesting are highly correlated, we retain both as they capture distinct and complementary dimensions of structural complexity (control flow versus nested depth) that are commonly used in software engineering studies.}

\begin{table}[htbp]
\centering
\small
\setlength{\tabcolsep}{4pt}
\renewcommand{\arraystretch}{1.05}
\caption{Code Quality Metrics from Understand.}
\begin{tabularx}{\columnwidth}{lX}
\toprule
\textbf{Metric} & \textbf{Definition} \\
\midrule
\textbf{Cyclomatic Complexity} & Number of independent paths through code. Higher values = more complex control flow. \\
\textbf{Max Nesting Depth} & Deepest level of nested structures (loops, conditionals). Deeper = harder to maintain. \\
\textbf{Total Methods/Class} & All methods including inherited. Measures class complexity and coupling. \\
\textbf{Comment/Code Ratio} & Ratio of comment to code lines. Indicates documentation density. \\
\textbf{Max Inheritance Depth} & Longest inheritance chain. Deeper = increased coupling, reduced flexibility. \\
\textbf{Functions per File} & Function declarations per file. Assesses modularity and single-responsibility. \\
\textbf{Methods in Class} & Methods declared within class (excluding inherited). Measures interface size. \\
\bottomrule
\end{tabularx}
\label{tab:understand-metrics}
\end{table}

\begin{table}[htbp]
\centering
\small
\setlength{\tabcolsep}{4pt}
\renewcommand{\arraystretch}{1.05}
\caption{Qodana severity classes.}
\begin{tabularx}{\columnwidth}{X}
\toprule

\textbf{Critical}: Must-fix defects. \\
\textit{Example:} In C++, dereferencing a pointer that was never initialised can crash the program (null-pointer dereference). \\ \midrule
\textbf{High}: Likely bugs or major smells. \\
\textit{Example:} A Java method annotated \emph{@NonNull} nevertheless returns \emph{null}, risking a run-time null-pointer exception. \\ \midrule
\textbf{Moderate}: Maintainability or style issues. \\
\textit{Example:} Two identical 12-line blocks appear in different functions, increasing future maintenance effort. \\ \bottomrule
\end{tabularx}

\label{tab:qodana_severities}
\end{table}

\subsubsection{\textbf{Mapping Contribution Patterns To Code Quality}}
\label{sec:rq3_analysis}

We analyze how the four core contribution patterns identified in RQ2 (see Section \ref{RQ2:Time series clustring}) affect code quality through a three-step process: snapshot creation, pattern mapping, and statistical clustering.

\paragraph{Project snapshot creation}
We create four snapshots for each project at key life stage points. Three snapshots correspond to the end of each development stage from RQ1: \textbf{early stage}, \textbf{mid stage}, and \textbf{late stage}. The fourth snapshot, the \textbf{initial development checkpoint}, captures the end of the first quartile of the early stage. The \textbf{initial development checkpoint} provides a meaningful baseline by ensuring the project contains substantial progress. At each snapshot, we use Git reset to revert the repository to the commit with the closest date to the selected time, then clone the complete source code to preserve the project's exact state.

For each snapshot, we apply both Qodana and Understand tools to extract quality metrics. We normalize all metrics by the number of source code characters to ensure fair comparison across projects of different sizes. We then calculate the percentage change in quality metrics (\(\Delta\)) between consecutive snapshots. For example, early stage quality change is calculated as follows: 

\[metric\Delta_{\text{early}} = \frac{\text{metric}_{\text{end\_of\_early}} - \text{metric}_{\text{initial\_checkpoint}}}{\text{metric}_{\text{initial\_checkpoint}}} \times 100\]

We map each Delta value to the core contribution pattern (e.g., \textit{High Constant Core}) that was active during that specific stage, as determined in RQ2(see Section \ref{RQ2:Time series clustring}).

\paragraph{Statistical Clustering}
We group the Delta values by their associated contribution pattern and apply the Scott-Knott test~\cite{scott1974cluster, noei2023empirical}. This test clusters patterns based on their statistical impact on each quality metric. Patterns in the same cluster show no significant difference in their quality impact, while patterns in different clusters have statistically distinct effects.

\subsection{Findings}

Table \ref{tab:sk_rank_by_metric} reveals
considerably different approaches to quality management between OSS4SG and conventional OSS projects. In this table, rank 1 \& 2 reflect strong performance, rank 3 a moderate performance, and ranks 4 \& 5 a poor performance. For complexity metrics, rank 1 means the lowest (strongest) complexity, while for documentation and organization metrics, rank 1 means the highest (strongest) quality. For issue severity, rank 1 indicates the fastest resolution times. To understand these differences, we analyze how each core contribution pattern affects code quality across these dimensions for both OSS4SG and conventional OSS projects.

\subsubsection{\textbf{OSS4SG Pattern Analysis}}
\takeaway{OSS4SG projects exhibit centralized quality management, with core contributors handling both structural oversight and issue resolution responsibilities.} The data in Table \ref{tab:sk_rank_by_metric} show that core contributors drive both quality control and issue resolution in OSS4SG projects. High Constant Core (\textsc{4SG-H}) achieves strong organizational metrics (rank 1 in all metrics) and strong to moderate issue control. As the core contributor ratio decreases from High Constant Core (\textsc{4SG-H}) to Low Core Activity (\textsc{4SG-L}), issue resolution performance declines across all severity levels (from ranks 1-3 to ranks 4-5).

\textbf{High Constant Core \textsc{(4SG-H)}} stands out as the most favorable pattern for OSS4SG projects. \textsc{4SG-H} ranks first in organizational metrics (methods, functions, declared methods) and moderate issue resolution among OSS4SG patterns. However, it shows mixed complexity results: low cyclomatic complexity (rank 2) and inheritance depth (rank 1), but the highest nesting complexity (rank 4).

\textbf{Increasing Core Activity \textsc{(4SG-I)}} maintains first-rank organizational structure (methods, functions, declared methods) with strong cyclomatic complexity control (rank 2) and moderate max inheritance complexity control (rank 3). However, \textsc{4SG-I} shows poor issue resolution, particularly for moderate-severity issues (rank 5) and max nesting complexity (rank 4).



\textbf{Low Core Activity \textsc{(4SG-L)}} shows best organizational quality (Methods: rank 1; Functions: rank 1; Declared methods: rank 2) and mixed complexity control (Cyclomatic: rank 3; Max Nesting: rank 2; Max inheritance: rank 3). However, \textsc{(4SG-L)} consistently underperforms in issue resolutions (Critical: rank 4; Rank high: 4; Moderate: rank 5).

\textbf{Declining Core Activity \textsc{(4SG-D)}} achieves the highest documentation ratio (rank 1) compared to all other 4SG patterns (rank 3). \textsc{(4SG-D)} also displays strong inheritance complexity control (rank 1). The high documentation ratio may indicate that departing or temporarily disengaging \textsc{4SG} contributors document their work to mitigate knowledge loss, reflecting their commitment to the project's mission and continuity. This might signal to maintainers that core contributors might be stepping away, either temporarily or permanently.

\textsc{(4SG-D)} is the least favorable OSS4SG pattern in terms of organizational structure. In fact, all other \textsc{4SG} patterns rank well in organizational metrics (ranks 1 and 2), whereas \textsc{4SG-D} falls behind. \textsc{(4SG-D)} is also the least favorable pattern for critical issue resolution (rank 5).





\subsubsection{\textbf{OSS Pattern Analysis}}Conventional OSS projects exhibit more distributed quality management, with distinct contributor engagement patterns tackling different aspects of code quality. The data shows that \takeaway{casual contributors \textsc{(OSS-L)} drive issue resolution in conventional OSS projects}. 
In contrast, \takeaway{core contributors excel at complexity control}, with High Constant Core (\textsc{OSS-H}) achieving the strongest cyclomatic complexity control (rank 1). This indicates that conventional OSS projects rely on casual contributors for issue handling and core contributors for complexity oversight.



\textbf{High Constant Core \textsc{(OSS-H)}} achieves the strongest cyclomatic complexity control (rank 1) and strong method organizational structure (Methods: rank 1; Decl. methods: rank 2). However, \textsc{(OSS-H)} shows a decline in issue resolution control as issue severity increases. This pattern shows a potential trade-off between complexity control and issue resolution speed.


\textbf{Increasing Core Activity \textsc{(OSS-I)}} demonstrates the strongest documentation quality (rank 1) and nesting complexity control (rank 1) but shows mixed performance elsewhere. In fact, \textsc{(OSS-I)} pattern exhibits moderate to poor issue resolution 
(critical: rank 4; High: rank 3; Moderate: rank 4).

\textbf{Low Core Activity \textsc{(OSS-L)}} show strong issue resolution for both critical and high-severity issues (ranks 1) across all \textsc{(OSS)} patterns and achieves top documentation quality (rank 1). However, \textsc{(OSS-L)} pattern shows mixed complexity control (Cyclomatic: rank 2; Max inheritance: rank 2; max Nesting: rank 4). 



\textbf{Declining Core Activity \textsc{(OSS-D)}} shows the poorest complexity control  (Cyclomatic: rank 4; Max Inheritance: rank 4) compared to all \textsc{(OSS)} patterns, and mixed organizational structure (Methods: rank 2; Functions: rank 4; Decl. Methods: rank 3). While \textsc{(OSS-D)} pattern maintains strong issue resolution for critical and moderate severity (ranks 2), but struggles with high-severity issues (rank 5).



\begin{tcolorbox}[
    colback=gray!10,
    colframe=gray!10,
    colbacktitle=gray!60, 
    coltitle=white, 
    title=RQ3 Summary,
    sharp corners,
    toptitle=1ex  
]

Contributor engagement patterns affects code quality approaches in both ecosystems. OSS4SG projects relies heavily on core contributors for both structural quality and issue resolution, with quality declining as core activity decreases. On the other hand, conventional OSS projects have a distributed quality control relying on casual contributors to resolve issues, while core contributors focus on managing project complexity.
\end{tcolorbox}

\section{Discussion}


Our findings reveal key differences in how mission-driven development shapes project sustainability. OSS4SG projects attract contributors who are personally invested in the social mission of the project, creating deeper engagement than conventional OSS's technical or career motivations. This stronger engagement leads to more stable communities where contributors remain consistently engaged over time. This results in sustained code quality throughout the project's life stages, as committed contributors maintain both structural integrity and issue resolution. 
In contrast, conventional OSS projects rely on a constant influx of new contributors to compensate for higher turnover, creating a different but equally functional model based on a more distributed effort rather than sustained commitment.

\hl{Our analysis of contributor email domains shows similar levels of corporate participation (OSS: 35.6\% vs. OSS4SG: 37.4\%) determined by excluding known personal email providers (e.g., Gmail, Yahoo) and bot accounts, indicating that the observed differences in community dynamics stem from intrinsic motivational factors rather than organizational sponsorship. The retention advantage observed in OSS4SG projects is most pronounced in early stages, suggesting that conventional OSS maintainers can learn from this pattern by nurturing newcomers during this period, so they transition into core contributors who sustain the project through later stages.}

Understanding how different types of contributor motivation shape community dynamics offers significant insights to OSS maintainers. This deeper personal investment manifests in subtle but telling behaviors: OSS4SG contributors write longer commit messages, maintain higher comment-to-code ratios, contribute during weekends, and even increase documentation prior to disengaging temporarily or permanently. Researchers can explore how these different motivations influence specific contributor behaviors and long-term project outcomes across various open-source contexts. 
\hl{For example, in OSS4SG, patterns of Declining Core Activity show the highest documentation ratios, reflecting how contributors take time to document their work rather than abruptly departing, ensuring project sustainability. The substantially higher weekend contribution rate among OSS4SG core contributors demonstrates commitment that extends beyond typical working hours. Conventional OSS maintainers can cultivate this level of engagement by connecting their project's goals to broader societal purposes. The contrasting strengths also reveal opportunities for cross-learning. OSS4SG projects, despite their social importance, achieve lower visibility than conventional OSS projects. OSS4SG project managers could learn from the more effective visibility strategies of conventional OSS to help their projects reach broader audiences while maintaining their mission focus
.}
\hl{OSS4SG projects can increase visibility by participating in events such as Google Summer of Code and Hacktoberfest\cite{DigitalOcean2014Hacktoberfest}, which are prevalent in conventional OSS projects as evidenced by event-related issue labels.}
Conversely, conventional OSS project managers could benefit from the OSS4SG approach to creating stronger contributor engagement through explicit social impact narratives, potentially improving their community stability.

\section{Threats to Validity}
Like any empirical study, our work has limitations, which we address in this section along with our mitigation strategies.

\textbf{Internal Validity.}
\hl{Within our project selection, we applied a systematic approach to remove toy projects, inactive projects, and projects with insufficient history from our original dataset. This filtering ensured that both OSS and OSS4SG project groups contained projects with comparable levels of maturity, activity, and longevity, reducing bias introduced by uneven development dynamics. However, some projects may still vary in how their communities work or contribute, which could affect the observed differences.}

A frequent limitation in mining OSS repositories is that developers may appear under several different identities. To mitigate this, we experiment with different methods and employ email-based matching and username, and email normalization \cite{zhu2019empirical}. Even with this process, some developers’ multiple identities may remain unresolved by the tested methods, potentially adding a small amount of noise to the contributor’s data. Furthermore, while these methods perform comparably in our dataset, their relative performance may differ in other datasets. Given the small variance observed across methods and the robustness of our findings, we believe this potential bias has a limited impact on our overall conclusions.

\textbf{Construct Validity.} Our classification of OSS4SG projects relies on existing Ovio \cite{ovio} and DPGA \cite{DPGARegistry} catalogs. Although these external sources are integral to our classification process, they are reliable and have been used in prior literature \cite{Huang2021, guizani2022attracting}.
The projects in our dataset vary widely in size, which could affect our comparisons. To address this, we normalize all metrics by code character counts, ensuring a fair comparison across different project sizes and programming languages. Additionally, we recognize that some projects may store portions of their code externally, which could lead to an underestimation of their size and affect our size-based comparisons.

\textbf{External Validity.} Our static analysis focuses on three programming languages: Python, Java, and JavaScript. This selection is influenced by the limitations of statistical analysis tools. For instance, in Qodana, each language operates within its own environment, and support is mostly limited to the most common languages. Additionally, structural metrics are not language-agnostic, leading to coverage gaps for less common languages. To mitigate this, we focus on the top three programming languages, which represent 65\% of our dataset and are widely used in both OSS and real-world applications. We also acknowledge that our conventional OSS sample may not fully represent all OSS projects. To assess this threat, we manually compared our sample with the top-starred GitHub projects \cite{markovtsev2018pga}, finding that 30\% of our conventional OSS projects are among the top-starred, confirming a good representativeness of popular OSS development. Furthermore, we have applied inclusion criteria from prior literature \cite{Pantiuchina2021} to ensure our sample consists of active projects with substantial numbers of contributors and contributions.

\hl{Our study builds on an existing dataset from prior work \cite{Huang2021}, which may not capture all OSS projects. However, the dataset of 1,039 projects exceeds the required sample size of 385 projects for a statistically representative sample of the broader OSS ecosystem (95\% confidence level, 5\% margin of error \cite{israel1992determining, ahmad2017determining}).} \hl{Our dataset represents established OSS4SG projects that are publicly listed in the Ovio and DPGA\cite{ovio, DPGARegistry}. While at the time of our study no new projects have been added to Ovio and DPGA, our dataset may not fully capture newly emerging projects that operate outside formal catalogues. Therefore, our findings may not be generalizable to the broader ecosystem of well-established OSS4SG projects.}

\section{Conclusion}
In this study, we provide the first large scale comparison of community dynamics and development practices between OSS4SG and conventional OSS projects. Our analysis of open source projects across over 3 million commits revealed key differences between the two ecosystems. We found that mission-driven projects form significantly more stable and ``sticky'' communities compared to conventional OSS. This stability can be attributed to OSS4SG's distinct engagement patterns. OSS4SG communities are characterized by consistent, high year-round activity from core members, while conventional OSS projects often face seasonal fluctuations and declining core involvement over time. Our findings reveals different approaches to quality management between OSS4SG and conventional OSS projects. These findings open opportunities for intervention studies testing whether conventional OSS projects can improve retention by incorporating mission narratives, and whether OSS4SG projects can scale by adopting OSS's distributed contribution mechanisms.

\section{Acknowledgment}
This work was supported by NSERC Discovery Grant RGPIN-2024-06511.

\bibliographystyle{ACM-Reference-Format}
\bibliography{biblio}

@String{Computing = "Computing" }

@String{Computer = "{IEEE} Computer" }

@String{Springer = "Springer-Verlag" }

@misc{little-window,
  title        = {Little Window},
  howpublished = {\url{https://github.com/chaynHQ/little-window}},
  note         = {Accessed: March 12, 2025}
}

@misc{commcare,
  title        = {CommCare},
  howpublished = {\url{https://www.dimagi.com/commcare/}},
  note         = {Accessed: March 12, 2025}
}

@online{DPGARegistry,
  title        = {Digital Public Goods Alliance Registry},
  organization = {Digital Public Goods Alliance},
  year         = {2024},
  url          = {https://digitalpublicgoods.net/registry/},
  urldate      = {2025-07-18},
  note         = {Accessed: 2025-07-18}
}

@inproceedings{gerosa2021motivation,
  title={The Shifting Sands of Motivation: Revisiting What Drives Contributors in Open Source},
  author={Gerosa, Marco and Wiese, Igor and Trinkenreich, Bianca and Link, Georg and Robles, Gregorio and Treude, Christoph and Steinmacher, Igor and Sarma, Anita},
  booktitle={Proceedings of the 43rd International Conference on Software Engineering (ICSE)},
  year={2021}
}

@inproceedings{olsson2017relationship,
  title={The relationship of code churn and architectural violations in the open source software jabref},
  author={Olsson, Tobias and Ericsson, Morgan and Wingkvist, Anna},
  booktitle={Proceedings of the 11th european conference on software architecture: companion proceedings},
  pages={152--158},
  year={2017}
}

@article{noei2023empirical,
  title={An empirical study of refactoring rhythms and tactics in the software development process},
  author={Noei, Shayan and Li, Heng and Georgiou, Stefanos and Zou, Ying},
  journal={IEEE Transactions on Software Engineering},
  volume={49},
  number={12},
  pages={5103--5119},
  year={2023},
  publisher={IEEE}
}

@article{spearman1961general,
  title={" General Intelligence" Objectively Determined and Measured.},
  author={Spearman, Charles},
  year={1961},
  publisher={Appleton-Century-Crofts}
}

@article{ghaleb2022popularity,
  title={On the Popularity of Internet of Things Projects in Online Communities: An Empirical Study of Hackster. io},
  author={Ghaleb, Taher Ahmed and da Costa, Daniel Alencar and Zou, Ying},
  journal={Information Systems Frontiers},
  volume={24},
  number={5},
  pages={1601--1634},
  year={2022},
  publisher={Springer}
}

@misc{wiki:KernelDensityEstimation,
  author = {{Wikipedia contributors}},
  title = {{Kernel density estimation} --- {W}ikipedia{,} The Free Encyclopedia},
  year = {2025},
  month = {August},
  day = {9},
  url = {https://en.wikipedia.org/w/index.php?title=Kernel_density_estimation&oldid=1305066115},
  note = {[Online; accessed 7-November-2025]}
}

@misc{DigitalOcean2014Hacktoberfest,
  author = {{DigitalOcean}},
  title = {{Hacktoberfest: A month-long celebration of open source}},
  year = {2014},
  howpublished = {\url{https://hacktoberfest.com}},
  note = {Accessed: 2025-11-07}
}

@software{OppiaTeam2013Oppia,
  author = {{The Oppia team}},
  title = {{Oppia: A free, online learning platform to make quality education accessible for all}},
  year = {2013},
  url = {https://github.com/oppia/oppia},
  note = {Accessed: 2025-11-07}
}

@software{Crosby2010Vimium,
  author = {Crosby, Phil and Sukhar, Ilya},
  title = {{Vimium: The hacker's browser}},
  year = {2010},
  url = {https://github.com/philc/vimium},
  note = {Accessed: 2025-11-07}
}

@inproceedings{zhang2014towards,
  title={Towards building a universal defect prediction model},
  author={Zhang, Feng and Mockus, Audris and Keivanloo, Iman and Zou, Ying},
  booktitle={Proceedings of the 11th working conference on mining software repositories},
  pages={182--191},
  year={2014}
}

@article{israel1992determining,
  title={Determining sample size},
  author={Israel, Glenn D and others},
  year={1992},
  publisher={University of Florida Cooperative Extension Service, Institute of Food and~…}
}

@article{ahmad2017determining,
  title={Determining sample size for research activities: the case of organizational research},
  author={Ahmad, Halim and Halim, Hasnita},
  journal={Selangor Business Review},
  pages={20--34},
  year={2017}
}

@inproceedings{Marlow2013CSCW,
 author = {Marlow, Jennifer and Dabbish, Laura and Herbsleb, Jim},
 title = {Impression Formation in Online Peer Production: Activity Traces and Personal Profiles in Github},
 booktitle = {Proceedings of the 2013 Conference on Computer Supported Cooperative Work},
 series = {CSCW '13},
 year = {2013},
 pages = {117--128},
 numpages = {12},
 publisher = {ACM},
 address = {New York, NY, USA},
}

@inproceedings{Singer2013CSCW,
 author = {Singer, Leif and Figueira Filho, Fernando and Cleary, Brendan and Treude, Christoph and Storey, Margaret-Anne and Schneider, Kurt},
 title = {Mutual Assessment in the Social Programmer Ecosystem: An Empirical Investigation of Developer Profile Aggregators},
 booktitle = {Proceedings of the 2013 Conference on Computer Supported Cooperative Work},
 series = {CSCW '13},
 year = {2013},
 pages = {103--116},
 numpages = {14},
 publisher = {ACM},
 address = {New York, NY, USA}
}

@article{guizani2021long,
  title={The Long Road Ahead: Ongoing Challenges in Contributing to Large OSS Organizations and What to Do},
  author={Guizani, Mariam and Chatterjee, Amreeta and Trinkenreich, Bianca and May, Mary Evelyn and Noa-Guevara, Geraldine J and Russell, Liam James and Cuevas Zambrano, Griselda G and Izquierdo-Cortazar, Daniel and Steinmacher, Igor and Gerosa, Marco A and others},
  journal={Proceedings of the ACM on Human-Computer Interaction},
  volume={5},
  number={CSCW2},
  pages={1--30},
  year={2021},
  publisher={ACM New York, NY, USA}
}

@inproceedings{sharma2012examining,
  title={Examining turnover in open source software projects using logistic hierarchical linear modeling approach},
  author={Sharma, Pratyush N and Hulland, John and Daniel, Sherae},
  booktitle={Open Source Systems: Long-Term Sustainability: 8th IFIP WG 2.13 International Conference, Hammamet, Tunisia. Proceedings 8},
  pages={331--337},
  year={2012},
  organization={Springer}
}

@inproceedings{guizani2022perceptions,
  title={Perceptions of the State of D\&I and D\&I Initiative in the ASF},
  author={Guizani, Mariam and Trinkenreich, Bianca and Castro-Guzman, Aileen Abril and Steinmacher, Igor and Gerosa, Marco and Sarma, Anita},
  booktitle={Proceedings of the ACM/IEEE 44th ICSE-SEIS},
  pages={130--142},
  year={2022}
}

@InProceedings{pinto2016more,
  author       = {Pinto, Gustavo and Steinmacher, Igor and Gerosa, Marco Aur{\'e}lio},
  title        = {More Common Than You Think: An In-depth Study of Casual Contributors},
  booktitle    = {IEEE International Conference on Software Analysis, Evolution, and Reengineering (SANER)},
  year         = {2016},
  volume       = {1},
  series       = {SANER 2016},
  pages        = {112--123},
  organization = {IEEE},
}

@inproceedings{steinmacher2018almost,
  title={Almost there: A study on quasi-contributors in open-source software projects},
  author={Steinmacher, Igor and Pinto, Gustavo and Wiese, Igor Scaliante and Gerosa, Marco Aur{\'e}lio},
  booktitle={2018 IEEE/ACM 40th ICSE },
  pages={256--266},
  year={2018},
  organization={IEEE}
}

@inproceedings{ferreira2020turnover,
  title={Turnover in Open-Source Projects: The Case of Core Developers},
  author={Ferreira, Fabio and Silva, Luciana Lourdes and Valente, Marco Tulio},
  booktitle={Proceedings of the XXXIV SBES},
  pages={447--456},
  year={2020}
}

@inproceedings{gerosa2021shifting,
  title={The shifting sands of motivation: Revisiting what drives contributors in open source},
  author={Gerosa, Marco and Wiese, Igor and Trinkenreich, Bianca and Link, Georg and Robles, Gregorio and Treude, Christoph and Steinmacher, Igor and Sarma, Anita},
  booktitle={2021 IEEE/ACM 43rd International Conference on Software Engineering (ICSE)},
  pages={1046--1058},
  year={2021},
  organization={IEEE}
}

@inproceedings{miller2019people,
  title={Why do people give up flossing? a study of contributor disengagement in open source},
  author={Miller, Courtney and Widder, David Gray and K{\"a}stner, Christian and Vasilescu, Bogdan},
  booktitle={Open Source Systems: 15th IFIP WG 2.13 International Conference, Montreal, QC, Canada, Proceedings 15},
  pages={116--129},
  year={2019},
  organization={Springer}
}

@inproceedings{steinmacher2013newcomers,
  title={Why do newcomers abandon open source software projects?},
  author={Steinmacher, Igor and Wiese, Igor and Chaves, Ana Paula and Gerosa, Marco Aur{\'e}lio},
  booktitle={6th International Workshop on Cooperative and Human Aspects of Software Engineering (CHASE)},
  pages={25--32},
  year={2013},
  organization={IEEE}
}

@InProceedings{Steinmacher.Chaves.ea_2014,
  author    = {Igor Steinmacher and Ana Paula Chaves and Tayana Conte and Marco Aur{\'e}lio Gerosa},
  title     = {Preliminary empirical identification of barriers faced by newcomers to Open Source Software projects.},
  booktitle = {Proceedings of the 28th SBES},
  year      = {2014},
  series    = {SBES '14},
  pages     = {51--60},
  publisher = {IEEE Computer Society},
  location  = {Macei{\'o}, Brazil, 28 September-3 October 2014},
}

@inproceedings{steinmacher2015social,
  title={Social barriers faced by newcomers placing their first contribution in open source software projects},
  author={Steinmacher, Igor and Conte, Tayana and Gerosa, Marco Aur{\'e}lio and Redmiles, David},
  booktitle={Proceedings of the 18th ACM conference on Computer supported cooperative work \& social computing},
  pages={1379--1392},
  year={2015}
}

@article{goggins2021making,
  title={Making open source project health transparent},
  author={Goggins, Sean P and Germonprez, Matt and Lumbard, Kevin},
  journal={Computer},
  volume={54},
  number={8},
  pages={104--111},
  year={2021},
  publisher={IEEE}
}

@inproceedings{feng2022case,
  title={A case study of implicit mentoring, its prevalence, and impact in Apache},
  author={Feng, Zixuan and Chatterjee, Amreeta and Sarma, Anita and Ahmed, Iftekhar},
  booktitle={Proceedings of the 30th ACM ESEC/FSE},
  pages={797--809},
  year={2022}
}

@article{trinkenreich2020hidden,
  title={Hidden figures: Roles and pathways of successful oss contributors},
  author={Trinkenreich, Bianca and Guizani, Mariam and Wiese, Igor and Sarma, Anita and Steinmacher, Igor},
  journal={Proceedings of the ACM on human-computer interaction},
  volume={4},
  number={CSCW2},
  pages={1--22},
  year={2020},
  publisher={ACM New York, NY, USA}
}

@misc{ovio,
    author = "{Ovio}",
    title = {{Contribute to open-source. Be part of the future!}},
    howpublished = {\url{https://ovio.org/}},
    note = {Online; accessed 2021} ,
    year=2021,
}

@inproceedings{markovtsev2018pga,
  author    = {Vadim Markovtsev and Waren Long},
  title     = {Public Git Archive: A Big Code Dataset for All},
  booktitle = {Proc.\ 15th Int.\ Conf.\ on Mining Software Repositories},
  pages     = {207--216},
  year      = {2018},
  publisher = {ACM},
  doi       = {10.1145/3196398.3196464}
}

@inproceedings{fang2023four,
  title={A Four-Year Study of Student Contributions to OSS vs. OSS4SG with a Lightweight Intervention},
  author={Fang, Zihan and Endres, Madeline and Zimmermann, Thomas and Ford, Denae and Weimer, Westley and Leach, Kevin and Huang, Yu},
  booktitle={Proceedings of the 31st ACM Joint European Software Engineering Conference and Symposium on the Foundations of Software Engineering},
  pages={3--15},
  year={2023}
}

@inproceedings{guizani2023rules,
  title={Rules of engagement: Why and how companies participate in OSS},
  author={Guizani, Mariam and Castro-Guzman, Aileen Abril and Sarma, Anita and Steinmacher, Igor},
  booktitle={2023 IEEE/ACM 45th International Conference on Software Engineering (ICSE)},
  pages={2617--2629},
  year={2023},
  organization={IEEE}
}

@inproceedings{zhu2019empirical,
  title={An empirical study of multiple names and email addresses in oss version control repositories},
  author={Zhu, Jiaxin and Wei, Jun},
  booktitle={2019 IEEE/ACM 16th International Conference on Mining Software Repositories (MSR)},
  pages={409--420},
  year={2019},
  organization={IEEE}
}

@article{amreen2020alfaa,
  title={ALFAA: Active Learning Fingerprint based Anti-Aliasing for correcting developer identity errors in version control systems},
  author={Amreen, Sadika and Mockus, Audris and Zaretzki, Russell and Bogart, Christopher and Zhang, Yuxia},
  journal={Empirical Software Engineering},
  volume={25},
  number={2},
  pages={1136--1167},
  year={2020},
  publisher={Springer}
}

@misc{githubAPI,
  author = "{GitHub, Inc.}",
  title = "{GitHub REST API v3}",
  year = {2024},
  howpublished = "\url{https://docs.github.com/en/rest}",
  note = "Accessed: 2024-02-16"
}

@misc{githubGraphQL,
  author = "{GitHub, Inc.}",
  title = "{GitHub GraphQL API v4}",
  year = {2024},
  howpublished = "\url{https://docs.github.com/en/graphql}",
  note = "Accessed: 2024-02-16"
}

@book{Raymond1999,
  author = {Raymond, Eric S.},
  title = {The Cathedral and the Bazaar: Musings on Linux and Open Source by an Accidental Revolutionary},
  year = {1999},
  publisher = {O'Reilly Media}
}

@article{Crowston2005,
  author = {Crowston, Kevin and Howison, James},
  title = {The Social Structure of Free and Open Source Software Development},
  journal = {First Monday},
  volume = {10},
  number = {2},
  year = {2005}
}

@article{Hars2002,
  author = {Hars, Alexander and Ou, Shaosong},
  title = {Working for Free? Motivations for Participating in Open-Source Projects},
  journal = {International Journal of Electronic Commerce},
  volume = {6},
  number = {3},
  pages = {25--39},
  year = {2002}
}

@article{macbeth2011cliff,
  title={Cliff's Delta Calculator: A non-parametric effect size program for two groups of observations},
  author={Macbeth, Guillermo and Razumiejczyk, Eugenia and Ledesma, Rub{\'e}n Daniel},
  journal={Universitas Psychologica},
  volume={10},
  number={2},
  pages={545--555},
  year={2011},
  publisher={Pontificia Universidad Javeriana}
}

@inproceedings{nassif2017revisiting,
  title={Revisiting turnover-induced knowledge loss in software projects},
  author={Nassif, Mathieu and Robillard, Martin P},
  booktitle={2017 IEEE international conference on software maintenance and evolution (ICSME)},
  pages={261--272},
  year={2017},
  organization={IEEE}
}

@inproceedings{foucault2015impact,
  title={Impact of developer turnover on quality in open-source software},
  author={Foucault, Matthieu and Palyart, Marc and Blanc, Xavier and Murphy, Gail C and Falleri, Jean-Rmy},
  booktitle={Proceedings of the 2015 10th joint meeting on foundations of software engineering},
  pages={829--841},
  year={2015}
}

@inproceedings{jamieson2024predicting,
  title={Predicting open source contributor turnover from value-related discussions: An analysis of GitHub issues},
  author={Jamieson, Jack and Yamashita, Naomi and Foong, Eureka},
  booktitle={Proceedings of the 46th IEEE/ACM International Conference on Software Engineering},
  pages={1--13},
  year={2024}
}

@inproceedings{xiao2023early,
  title={How early participation determines long-term sustained activity in github projects?},
  author={Xiao, Wenxin and He, Hao and Xu, Weiwei and Zhang, Yuxia and Zhou, Minghui},
  booktitle={Proceedings of the 31st ACM Joint European Software Engineering Conference and Symposium on the Foundations of Software Engineering},
  pages={29--41},
  year={2023}
}

@inproceedings{guizani2022attracting,
  title={Attracting and retaining oss contributors with a maintainer dashboard},
  author={Guizani, Mariam and Zimmermann, Thomas and Sarma, Anita and Ford, Denae},
  booktitle={Proceedings of the 2022 ACM/IEEE 44th International Conference on Software Engineering: Software Engineering in Society},
  pages={36--40},
  year={2022}
}

@inproceedings{zhou2012make,
  title={What make long term contributors: Willingness and opportunity in OSS community},
  author={Zhou, Minghui and Mockus, Audris},
  booktitle={2012 34th International Conference on Software Engineering (ICSE)},
  pages={518--528},
  year={2012},
  organization={IEEE}
}

@manual{Understand,
  title        = {Understand: The Software Developer's Multi-Tool},
  author       = {{SciTools}},
  year         = {2025},
  note         = {Version 6.2},
  url          = {https://scitools.com/},
  note         = {Accessed: 2025-03-07}
}

@misc{Qodana,
  author       = {JetBrains},
  title        = {Qodana: Code Quality Platform},
  howpublished = {\url{https://www.jetbrains.com/qodana/}},
  note         = {Accessed: 2025-03-07}
}

@inproceedings{Huang2021,
  author = {Huang, Yuan and Ford, Denae and Zimmermann, Thomas},
  title = {Leaving My Fingerprints: Motivations and Challenges of Contributing to OSS for Social Good},
  booktitle = {Proceedings of the 43rd IEEE/ACM International Conference on Software Engineering (ICSE)},
  pages = {1020--1032},
  year = {2021},
  organization = {IEEE}
}

@article{scott1974cluster,
  title={A cluster analysis method for grouping means in the analysis of variance},
  author={Scott, Andrew Jhon and Knott, Martin},
  journal={Biometrics},
  pages={507--512},
  year={1974},
  publisher={JSTOR}
}

@article{noei2025empirical,
  title={An Empirical Study on Release-Wise Refactoring Patterns},
  author={Noei, Shayan and Li, Heng and Zhou, Ying},
  journal={Proceedings of the ACM on Software Engineering},
  volume={2},
  number={FSE},
  year={2025}
}

@article{roberts2006understanding,
  title={Understanding the motivations, participation, and performance of open source software developers: A longitudinal study of the Apache projects},
  author={Roberts, Jeffrey A and Hann, Il-Horn and Slaughter, Sandra A},
  journal={Management science},
  volume={52},
  number={7},
  pages={984--999},
  year={2006},
  publisher={INFORMS}
}

@misc{Huang2021dataset,
  title        = {Supplementary Material for {Leaving My Fingerprints: Motivations and Challenges of Contributing to OSS for Social Good}},
  author       = {Huang, Yu and Ford, Denae and Zimmermann, Thomas},
  year         = {2021},
  month        = feb,
  publisher    = {Zenodo},
  howpublished = {\url{https://doi.org/10.5281/zenodo.4536791}},
  note         = {Accessed: March 6, 2025}
}

@inproceedings{Pantiuchina2021,
  author = {Pantiuchina, Julia and Lin, Bo and Zampetti, Fabio and Di Penta, Massimiliano and Lanza, Michele and Bavota, Gabriele},
  title = {Why Do Developers Reject Refactorings in Open-Source Projects?},
  booktitle = {ACM Transactions on Software Engineering and Methodology (TOSEM)},
  volume = {31},
  number = {2},
  pages = {1--23},
  year = {2021},
  organization = {ACM}
}

@article{Wang2012,
  author = {Wang, Jian},
  title = {Survival Factors for Free Open Source Software Projects: A Multi-Stage Perspective},
  journal = {European Management Journal},
  volume = {30},
  number = {4},
  pages = {352--371},
  year = {2012}
}

@article{bock2021measuring,
  title={Measuring and modeling group dynamics in open-source software development: A tensor decomposition approach},
  author={Bock, Thomas and Schmid, Angelika and Apel, Sven},
  journal={ACM Transactions on Software Engineering and Methodology (TOSEM)},
  volume={31},
  number={2},
  pages={1--50},
  year={2021},
  publisher={ACM New York, NY}
}

@inproceedings{Ferreira2020,
  author = {Ferreira, Fernando and Silva, Lucas Lima and Valente, Marco Tulio},
  title = {Turnover in Open-Source Projects: The Case of Core Developers},
  booktitle = {Proceedings of the XXXIV Brazilian Symposium on Software Engineering (SBES)},
  pages = {447--456},
  year = {2020},
  organization = {ACM}
}

\end{document}